\documentclass[11pt,a4paper]{article} 
\usepackage{graphicx}
\usepackage{jcappub}

\newcommand{\pd}{\partial}  
\newcounter{ichi}
\newcounter{ni}
\newcounter{san}
\newcounter{yon}
\title{Constraining very heavy dark matter using diffuse backgrounds of neutrinos and cascaded gamma rays}

\author{Kohta Murase$^{1,2}$ and John F. Beacom$^{1,2,3}$}

\affiliation{
$^{1}$CCAPP, Ohio State University, 191 W. Woodruff Ave., Columbus, Ohio 43210, USA\\
$^{2}$Department of Physics, OSU, 191 W. Woodruff Ave., Columbus, Ohio 43210, USA\\
$^{3}$Department of Astronomy, OSU, 140 W. 18th Ave., Columbus, Ohio 43210, USA
}

\emailAdd{murase.2@osu.edu, beacom.7@osu.edu}

\abstract{
We consider multi-messenger constraints on very heavy dark matter (VHDM) from recent {\it Fermi} gamma-ray and IceCube neutrino observations of isotropic background radiation.  
{\it Fermi} data on the diffuse gamma-ray background (DGB) shows a possible unexplained feature at very high energies (VHE), which we have called the ``VHE Excess" relative to expectations for an attenuated power law extrapolated from lower energies.  We show that VHDM could explain this excess, and that neutrino observations will be an important tool for testing this scenario.  
More conservatively, we derive new constraints on the properties of VHDM for masses of ${10}^3$--${10}^{10}$~GeV.  These generic bounds follow from cosmic energy budget constraints for gamma rays and neutrinos that we developed elsewhere, based on detailed calculations of cosmic electromagnetic cascades and also neutrino detection rates.  
We show that combining both gamma-ray and neutrino data is essential for making the constraints on VHDM properties both strong and robust.  In the lower mass range, our constraints on VHDM annihilation and decay are comparable to other results; however, our constraints continue to much higher masses, where they become relatively stronger.
}


\begin{document}
\maketitle

\flushbottom

%%%%%%%%%%%%%%%%%%%%%%%%%%%%%%%%%%%%%%%%%%%%%%%%%%%%%%%%
%%%%%%%%%%%%%%%%%%%%%%%%%%%%%%%%%%%%%%%%%%%%%%%%%%%%%%%%

\section{Introduction}

A new era in high-energy multi-messenger astronomy is dawning.  Gamma-ray and neutrino observations will help probe the origins of cosmic rays, the nature of high-energy sources, the mechanisms that power astrophysical objects, and more.  In particular, these observations may help finally reveal the particle properties of dark matter.  Very heavy dark matter (VHDM), with $m_{\rm dm} c^2 >$~TeV, could produce negligible rates in nuclear scattering experiments, due to its low number density, and might be out of reach of collider production experiments, due to its high mass.  However, there are new opportunities to search for the products of its annihilation or decay, due to the increasing sensitivity of gamma-ray and neutrino experiments.   Here the high mass of VHDM may compensate its low number density, because so much energy ($2 m_{\rm dm} c^2$ or $m_{\rm dm} c^2$, respectively) is released per annihilation or decay.

In the GeV range, the LAT instrument onboard the {\it Fermi} satellite has found various extragalactic gamma-ray sources, including active galactic nuclei (AGN), star-burst galaxies and gamma-ray bursts (GRBs)~\citep[e.g.,][]{der12,abd+09,abd+10a,abd+11}.  The diffuse gamma-ray background (DGB)~\cite{dgb}, presumed to be cosmic, was measured by LAT in the GeV range~\cite{abd+10p,abd+10a2} and was found to be lower than that obtained by EGRET~\cite{egret}.  The origin of the DGB is unsettled; in any case, the measured spectrum leads to strong bounds on the gamma-ray emissivity from general sources in the universe~\cite{ca97,mur+12,ii12}, as well as in specific scenarios, e.g., fast redshift-evolution models of ultra-high-energy (UHE) cosmic rays~\cite{ahl+10}.

In the very-high-energy (VHE; $>0.1$~TeV) range, thanks to the big successes of imaging atmospheric Cherenkov telescopes (IACTs) such as HESS, MAGIC and VERITAS, not only Galactic but also extragalactic gamma-ray sources have been discovered~\citep[see a recent review, e.g.,][and references therein]{hh10}.  The next generation IACT, Cherenkov Telescope Array (CTA), is also being planned~\cite{act+11}; its sensitivity is expected to be improved by a factor of $\sim 10$ compared to existing IACTs.  However, the discovery power of gamma-ray experiments at the highest energies will always be limited by the opacity of the universe caused by pair production processes on the extragalactic background light (EBL) and the cosmic microwave background (CMB).

Neutrinos, however, are not attenuated in the cosmos, so are especially important to reveal distant, high-energy processes directly.  The IceCube detector at the South Pole was completed~\citep[e.g.,][]{ahr+04,abb+11a,rot+11,abb+11b,abb+11c} and the comparable KM3Net detector is being planned in the Mediterranean Sea~\cite{kat06}.  Interesting constraints on UHE neutrino fluxes have been placed with, e.g., the balloon-based ANITA experiment~\cite{gor+10}.  Although non-terrestrial high-energy neutrinos have not been detected so far, the sensitivity of present experiments is nearing well-motivated theoretical expectations~\cite{neu,mur11}.    

Various searches for dark matter have been made to find out indirect signals from its annihilation~\citep[e.g.,][]{anngs,ark+09,anngb,annsub,zav+10,annline,annneu} and decay~\citep[e.g.,][]{decold,decold2,decgs,grav,cov+09,hidden,composite}, including possible signatures in the DGB and the neutrino background.  Despite the fact that dark matter has been known since the 1930s~\cite{zwi33} by its gravitational effects, its particle properties remain unknown.  Numerous candidates have been suggested as a solution to this dark matter problem.  Among various possibilities, weakly interacting massive particles (WIMPs) are especially popular~\cite{dm}.  They are motivated by the problem of electroweak symmetry breaking, and neutral WIMPs like the neutralino have been studied in great detail.  Neutralinos are spin-1/2 Majorana fermions, predicted in the supersymmetric extension of the standard theory with R-parity conservation.  They can naturally be the stable lightest supersymmetric particle, and may have annihilation cross sections such that their thermal relic abundance matches the observed dark matter density.  This implies that WIMP-matter interactions may be strong enough that WIMPs will be produced at particle accelerators and detected directly in nuclear scattering experiments.  Self-annihilations of WIMPs in the present-day universe may lead to detectable gamma-ray and neutrino signals; their non-observation has led to a variety of limits on properties of annihilating WIMPs~\cite{anncong,anncong2,abd+10j,aba+10,abr+11,bea+07,yuk+07,por+11}.

Furthermore, there is no fundamental objection to considering unstable dark matter, as long as the lifetime is longer than the age of the universe.  This possibility has been studied and constrained in various aspects from many years ago~\citep[e.g.,][]{decold,decold2}, including more general cases where neutral relics are subdominant~\cite{ell+92,gon92,gon+93,kr97}.  Later, some specific particle physics models of decaying dark matter have recently been proposed~\citep[e.g.,][]{grav,cov+09,hidden,composite}.  One of the long-lived dark matter candidates in the supersymmetric theory is the gravitino in R-parity-breaking vacua, where long lifetimes are allowed by the supersymmetric breaking scale and small R-parity violation~\cite{grav,cov+09}.  Not only the gravitino but also the sneutrino can be viable decaying dark matter candidates as super-WIMPs~\cite{fen+03} that may not be seen by collider production and nuclear scattering experiments.  It is also possible to consider hidden sector gauge bosons, gauginos, and fermions as decaying dark matter~\cite{hidden}.  Composites of strongly-interacting particles have also been proposed~\cite{composite}.  Decaying dark matter scenarios have been constrained from the pre-\textit{Fermi} era~\cite{ell+92,gon92,gon+93,kr97,deccong,pal08}, and they have been recently discussed especially in the literature of the anomaly in the positron spectrum~\cite{por+11,deccong2,decconn}.    

For ordinary thermal relics, the required annihilation cross section is ${<\sigma v>}_{\rm dm} \approx (2-5) \times {10}^{-26}~{\rm cm}^{3}~{\rm s}^{-1}$~\citep[see][and references therein]{ste+12}.  
However, dark matter may be nonthermally produced in the early universe, and the cross section can be larger in principle.  Nonthermal production is commonly considered in the models of decaying dark matter.  While many models, including supersymmetric models, consider dark matter masses of $\lesssim 100$~TeV, super heavy dark matter, e.g., wimpzillas, has also been suggested in various contexts~\cite{shdmrev,kc85,shdm,shdm2,shdmcr} including earlier works motivated by the string theory~\cite{string}.  There are various production mechanisms, and the wimpzilla mass, $\sim {10}^{9}-{10}^{19}$~GeV, may be related to the inflaton mass~\cite{shdm,shdm2}.   Long lifetimes can be realized by considering breaking discrete gauge symmetries and some other non-perturbative effects~\cite{shdmcr,symbreak}.  Although the motivation to explain UHE cosmic rays seems to have almost disappeared~\cite{shdmrev,shdmcr}, super heavy dark matter is still viable as a dark matter candidate.

In this work, we focus on generic candidates for present-day decaying VHDM with masses between ${10}^3$ and ${10}^{10}$~GeV.  Since dark matter has not yet been discovered in the ``expected" lower mass range, we are motivated to search at higher energies, which are now being explored with much greater sensitivities.  There are also several independent works focusing on decaying VHDM~\cite{independent}.  
Another new point of our paper is to discuss the origin of the possible ``VHE Excess" in the DGB~\cite{mur+12}, which has been identified in the \textit{Fermi} era.  Dark matter with masses of $\sim 1-10$~GeV was discussed many years ago to explain the DGB in the MeV range~\cite{decold}.  Thanks to \textit{Fermi}, the VHE DGB is now seen well above the energy ranges of COMPTEL and EGRET, and we find that this new feature could be explained by dark matter, and that IceCube searches for neutrinos are already constraining some of the cases in this scenario.

Many previous studies on gamma rays focused on the mass range below $10$~TeV, where the gamma-ray cascade caused by pair production on the cosmic photon backgrounds is not relevant.  At higher masses, however, one has to take gamma-ray cascades into account properly.  Our detailed calculations allow us to derive the new cascade gamma-ray bound on the annihilation cross section and decay lifetime of VHDM.  Importantly, the \textit{Fermi} DGB is lower than the EGRET one, so past constraints on VHDM~\citep[e.g.,][]{ell+92,gon92,kr97} can be improved by about a factor $\sim 10$.  In addition, though neutrino constraints were considered in early days~\citep[e.g.,][]{ell+92,gon92,gon+93}, IceCube has indeed allow us to achieve limits that are stronger by orders of magnitude than those from the first generation experiments in the COMPTEL/EGRET era, e.g., from the Monopole Astrophysics and Cosmic Ray Observatory (MACRO) and the Irvine-Michigan-Brookhaven (IMB) detector.  Updated calculations of both the messengers are of importance and enable us to demonstrate that neutrino observations can be more stringent than gamma-ray observations at sufficiently high masses. 

The organization of this paper is as follows. 
In Section~2, we revisit constraints on the cosmic energy budget of gamma rays and neutrinos, based on results of {\it Fermi} and IceCube, providing an overview of how we restrict the properties of VHDM.  In Section~3, we show numerical results of the spectra arising from annihilating or decay VHDM.  In Section~4, we consider the most optimistic cases, where VHDM is assumed to contribute to the VHE DGB, and demonstrate the importance of neutrino observations.  In Section~5, we develop and synthesize our combined gamma-ray and neutrino constraints on annihilation and decay of VHDM.  In Section~6, we present our conclusions.  Two appendices present further details of the calculation of intergalactic electromagnetic cascades and how the Galactic foreground signals from dark matter are taken into account in our calculations of cosmic signals from VHDM.  In this paper, according to WMAP seven-year results~\cite{kom+10}, we adopt $H_0 \equiv 100h =70.2~{\rm km}~{\rm s}^{-1}~{\rm Mpc}^{-1}$, $\Omega_{\rm dm}=0.229$, $\Omega_m=0.275$ and $\Omega_{\Lambda}=0.725$. 

%%%%%%%%%%%%%%%%%%%%%%%%%%%%%%%%%%%%%%%%%%%%%%%%%%%%%%%%
%%%%%%%%%%%%%%%%%%%%%%%%%%%%%%%%%%%%%%%%%%%%%%%%%%%%%%%%

\section{Overview of cosmic energy budget considerations}

Indirect dark matter signatures from annihilation or decay can be searched for using diffuse background radiation.  The isotropic DGB has been measured, but its origin is not understood, so a dark matter contribution can be recognized only if it is significant relative to the full DGB, and not merely to its uncertainty.  If the DGB mainly consists of relatively rare point sources such as blazars, the DGB can be reduced by further progress in resolving point sources; the improvement is expected to be modest~\cite{aba+10}.  The neutrino background has not been detected so far, though atmospheric neutrinos have been measured up to several hundred TeV.  At higher energies, where the atmospheric neutrino flux is negligible, the bounds on dark matter properties are already powerful and will improve quickly.

It is useful to convert constraints from diffuse background fluxes into those on cosmic energy budgets.  
Gamma rays and neutrinos may be produced by astrophysical sources or dark matter.  In the case of dark matter annihilation and decay, these stable standard model particles will be among the dominant endpoints of all decay chains from heavier particles.  Both neutrinos and (lower-energy) gamma rays travel long distances, so that they are sensitive to processes occurring throughout the cosmos.  One cannot see as far with higher-energy gamma rays, because they induce electromagnetic cascades, as will electrons and positrons; all of these are included in our calculations.

Following Murase et al.~\cite{mur+12}, we briefly discuss the required or allowed energy budgets in gamma rays and neutrinos.  We denote the bolometric energy budget and differential energy budget by $Q(z)$ and $E Q_E (z)$, respectively.  For discrete sources, $Q (z)$ is the product of the luminosity $L (z)$ per source and the source density $n_s (z)$, and $E Q_E(z)$ is the same before integration over energies; for continuum sources, these generalize.  We consider two typical cases.  For a quasi-differential limit, we consider a near-mono-energetic injection spectrum.  For an integrated limit, we consider a $E^{-2}$ spectrum (i.e., $E Q_E \propto$~const.) spanning several decades.  Constraints are reported in terms of the local ($z = 0$) energy budgets, with the assumed redshift evolution noted.

Sufficiently high-energy gamma rays from distant sources cannot avoid pair creation with the EBL and CMB, and will be cascaded down to GeV-TeV energies through pair creation and inverse Compton scattering processes.  Whatever the origin of the DGB is, if the cascade is sufficiently developed, it has a near-universal form~\cite{mur+12,cascade},
\begin{equation}
G_{E_{\gamma}} \propto 
\left\{ \begin{array}{ll}
{(E_{\gamma}/E_{\gamma}^{\rm br})}^{-1/2}
& \mbox{($E_\gamma \leq E_{\gamma}^{\rm br}$)}\\
{(E_{\gamma}/E_{\gamma}^{\rm br})}^{1-\beta} 
& \mbox{($E_{\gamma}^{\rm br} < E_\gamma \leq E_{\gamma}^{\rm cut}$)}.
\end{array} \right. 
\end{equation} 
where the normalization of $G_{E_{\gamma}}$ is $\int d E_\gamma \, G_{E_{\gamma}} = 1$.  Here, $E_{\gamma}^{\rm cut}$ is the energy where the suppression due to pair creation occurs, $\beta$ is typically $\sim 2$, and $E_{\gamma}^{\rm br} \approx (4/3) {({E'}_\gamma^{\rm cut}/m_e c^2)}^2 \varepsilon_{\rm CMB} \simeq 0.034~{\rm GeV}~{(E_{\gamma}^{\rm cut}/0.1~{\rm TeV})}^2 {((1+z)/2)}^{2}$, where $\varepsilon_{\rm CMB}$ is the typical CMB energy. 

The DGB observed by LAT can be fitted by a power law~\cite{abd+10p},
\begin{equation}
E_{\gamma}^2 \Phi_{\gamma} = 0.855 \times {10}^{-7}~{\rm GeV}~{\rm cm}^{-2}~{\rm s}^{-1}~{\rm sr}^{-1}~{(E_{\gamma}/100~{\rm GeV})}^{-0.41},
\end{equation}
which can be crudely connected to the theoretical background gamma-ray flux as
\begin{eqnarray}
E_\gamma^2 \Phi_{\gamma} \approx \frac{c t_H}{4 \pi} (E_\gamma \bar{G}_{E_\gamma} Q_{\gamma}) \xi_z, 
\end{eqnarray}
where $Q_\gamma \equiv \int d E_\gamma \, \, Q_{E_\gamma}$ is the local gamma-ray energy budget and $\xi_z$ is the pre-factor coming from its evolution~\cite{wb98}.  For dark matter annihilation or decay, one has $\xi_z \sim 1$ as a typical value, because the most important contributions come from $z \sim 0$ as long as the dark matter lifetime is longer than the Hubble time $t_H$~\footnote{In the no redshift evolution case that is expected for decaying dark matter, we obtain $\xi_z \simeq 0.59$.  For annihilating dark matter, we get $\xi_z \simeq 0.86$ with the fitting function used in Yuksel et al.~\cite{yuk+07}.}.    
For astrophysical sources, important contributions often come from sources at $z \sim 1-2$, where
the typical redshift of $\bar{z} \sim 1$ leads to $E_\gamma^{\rm cut}|_{\bar{z} \sim 1} \sim 0.1$~TeV, so that one has $\xi_z \sim$~a few.  
Since cascade gamma rays have a characteristic spectrum and we typically have $E_\gamma \bar{G}_{E_\gamma} \sim 0.1$, details of primary spectra are not very relevant because they are washed out.  The DGB therefore probes the bolometric electromagnetic energy budget in the VHE/UHE range.  From Eq.~(2.3), the DGB obtained by LAT implies
\begin{equation}
Q_{\gamma} = L_\gamma n_s  \lesssim 2 \times {10}^{45}~{\rm erg}~{\rm Mpc}^{-3}~{\rm yr}^{-1} \xi_z^{-1},
\end{equation}
when the injection energy is high enough to induce cascades.  Our derived DGB constraint is shown in Figure~1.  

%++++++++++++++++++++++++++++++++++++++++++++++++++++++++++
\begin{figure*}[tb]
\begin{minipage}{0.49\linewidth}
\begin{center}
\includegraphics[width=\linewidth]{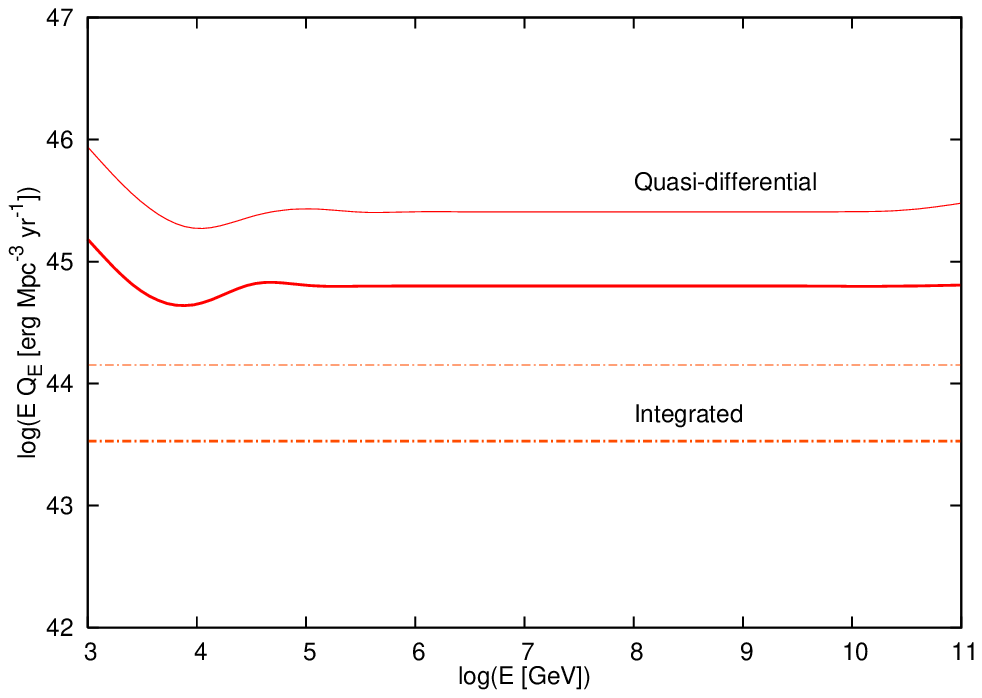}% Here is how to import EPS art
\caption{Upper bounds on the local ($z = 0$) gamma-ray energy budget of the universe.  
The upper solid curves represent quasi-differential limits obtained with $E' Q_{E'} = Q_\gamma$, i.e., the same energy input at any $E'$, for near-mono-energetic gamma-ray injection. 
The lower dot-dashed curves represent integrated limits for $E' Q_{E'} =$~const., i.e., an $E^{-2}$ differential spectrum, where ${E'}_{\rm min}={10}^{2.75}$~GeV and ${E'}_{\rm max}={10}^{11.25}$~GeV are assumed.   Thick curves are obtained for star formation evolution~\cite{hb06}, whereas thin curves are for no redshift evolution.  
Adapted from Ref.~\cite{mur+12}.
\newline \, \newline \, \newline \, \newline}
\end{center}
\end{minipage}
\begin{minipage}{.02\linewidth}
\end{minipage}
\begin{minipage}{0.49\linewidth}
\begin{center}
\includegraphics[width=\linewidth]{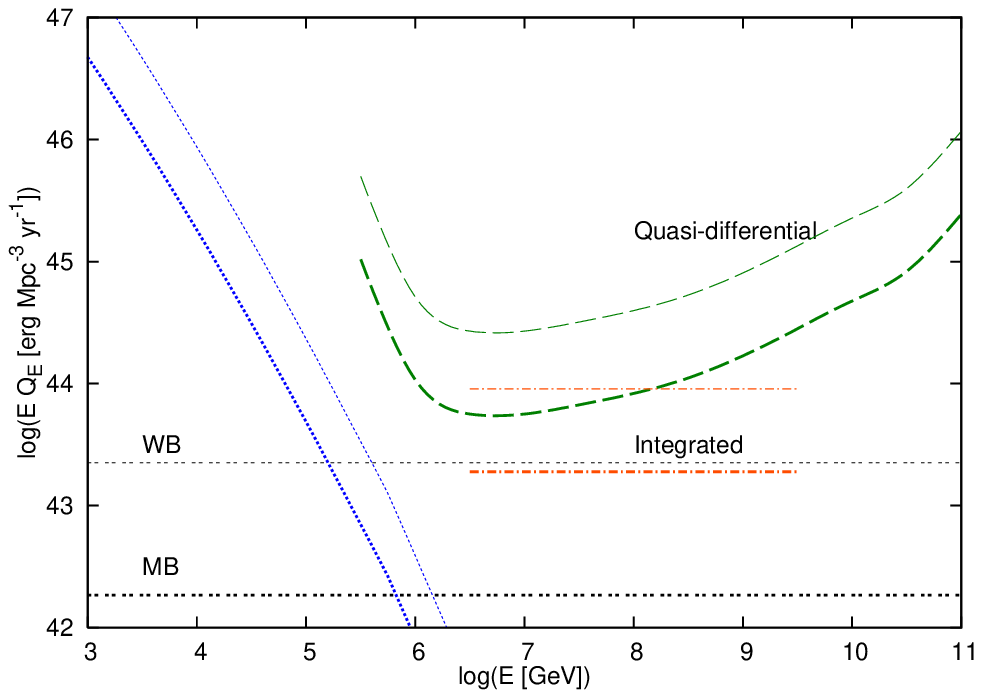}% Here is how to import EPS art
\caption{Upper bounds on the local ($z = 0$) neutrino energy budget of the universe.  The upper dashed curves represent quasi-differential limits estimated from the IceCube-40 analysis (333.5~days)~\cite{abb+11b}.  The lower dot-dashed curves represent integrated limits obtained for $E_{\nu} Q_{E_{\nu}} =$~const. (and a narrower energy range than that used in Figure~1).   The thick curves are obtained for star formation evolution~\cite{hb06}, whereas the thin curves are for no redshift evolution.  The dotted curves  show limits from the atmospheric neutrino background.  WB represents the Waxman-Bahcall bound~\cite{wb98} obtained with $E_{\rm cr} Q_{E_{\rm cr}} =0.6 \times {10}^{44}~{\rm erg}~{\rm Mpc}^{-3}~{\rm yr}^{-1}$, and MB represents the effective nucleus-survival bound of Murase \& Beacom~\cite{mb10}.  Adapted from Ref.~\cite{mur+12}.
\newline
}
\end{center}
\end{minipage}
\end{figure*}
%++++++++++++++++++++++++++++++++++++++++++++++++++++++++++

Unlike gamma rays, neutrinos reach the Earth through intergalactic space without attenuation.  Because of this, neutrino detectors are directly sensitive to the differential energy budget and spectrum in the VHE/UHE range, $E Q_E$.   Any non-terrestrial neutrino background has not been detected yet, but the IceCube-40 integrated limit obtained for the $E^{-2}$ spectrum has now reached an impressive sensitivity~\cite{abb+11b},
\begin{equation}
E_\nu^2 \Phi_\nu \lesssim {\rm a~few} \times {10}^{-8}~{\rm GeV}~{\rm cm}^{-2}~{\rm s}^{-1}~{\rm sr}^{-1};
\end{equation}
the typical differential sensitivity is weaker by a factor of a few. 
The theoretical extragalactic neutrino background is related to the energy budget as~\cite{wb98}
\begin{equation}
E_\nu^2 \Phi_\nu \approx \frac{c t_H}{4 \pi} (E_\nu Q_{E_\nu}) \xi_z.
\end{equation}
Here $E_\nu Q_{E_{\nu}}$ is the differential neutrino energy budget, which is constrained as
\begin{equation}
E_{\nu} Q_{E_{\nu}} \lesssim 4.3 \times {10}^{43}~{\rm erg}~{\rm Mpc}^{-3}~{\rm yr}^{-1}~\left( \frac{E_\nu^2 \Phi_\nu}{3 \times {10}^{-8}~{\rm GeV}~{\rm cm}^{-2}~{\rm s}^{-1}~{\rm sr}^{-1}}\right) \xi_z^{-1} . 
\end{equation}
The neutrino constraint is shown in Figure~2. 
Note that the neutrino flux limits can be below the gamma-ray measurements, so neutrinos are more sensitive in some cases, especially at high energies.  At low energies, the neutrino sensitivity is worse due to the atmospheric neutrino background.

It is often useful to compare the experimental neutrinos limits to theoretical bounds used in the literature of cosmic rays.  In Figure~2, we also show the nucleon-survival (Waxman-Bahcall) bound derived for cosmic-ray sources~\cite{wb98} and the nucleus-survival (Murase-Beacom) bound for sources of cosmic-ray nuclei~\cite{mb10}.  
The cosmogenic neutrino flux, which gives a guaranteed astrophysical neutrino background, practically obeys the nucleon-survival bound (for primary protons) and the nucleus-survival bound (for primary nuclei)~\cite{cosmon}.  

Our general constraints on the energy budgets can be adapted for dark matter.
For decaying dark matter, the local energy budget, $Q_{\rm dm}$, is
\begin{equation}
Q_{\rm dm}^{\rm dec} =  \frac{\rho_{\rm dm}}{m_{\rm dm}} \frac{m_{\rm dm} c^2}{\tau_{\rm dm}} = \frac{\rho_{\rm dm} c^2}{\tau_{\rm dm}} ,
\end{equation}
because each decay injects an energy $m_{\rm dm} c^2$.
For annihilating dark matter, one has
\begin{equation}
Q_{\rm dm}^{\rm ann} = {\left( \frac{\rho_{\rm dm}}{m_{\rm dm}} \right)}^2 \frac{{<\sigma v>}_{\rm dm}  g_0}{2} 2 m_{\rm dm} c^2 = \frac{{<\sigma v>}_{\rm dm} \rho_{\rm dm}^2 c^2 g_0}{m_{\rm dm}} ,
\end{equation}
because each annihilation injects an energy of $2 m_{\rm dm} c^2$.  Unlike for decay, here the clustering of dark matter is important, and $g_0$ is the flux-multiplier at $z=0$, which depends on the dark matter profile and substructures (see Section~3.2).

%%%%%%%%%%%%%%%%%%%%%%%%%%%%%%%%%%%%%%%%%%%%%%%%%%%%%%%%
%%%%%%%%%%%%%%%%%%%%%%%%%%%%%%%%%%%%%%%%%%%%%%%%%%%%%%%%

\section{Calculations of neutrino and gamma-ray spectra from VHDM}
The origin of the DGB is a mystery, and a number of models have been suggested so far~\citep[see recent reviews][and references therein]{dgb,der12}.  Whatever the origin is, in the standard theory (unless axion conversion or Lorentz-invariance violation is invoked), extragalactic gamma rays must be attenuated at sufficiently high energies due to the pair creation with EBL photons.  Therefore, one should expect suppression at $\gtrsim 100$~GeV energies, so that, \textit{even if} the originally-emitted DGB extends to higher energies, it will be severely attenuated.  Although careful data analyses and precise measurements of the VHE DGB are required, preliminary \textit{Fermi} data~\footnote{See the presentation by M. Ackermann on behalf of the {\it Fermi} collaboration: http://agenda.albanova.se/conferenceDisplay.py?confId=2600.} suggest the possible existence of a distinct component at $\gtrsim 100$~GeV~\cite{mur+12}, because the measurements are well above expectations for an attenuated power law extrapolated from lower energies.  We have called the discrepancy the ``VHE Excess" and shown that it can have a significant contribution from cascade components~\cite{mur+12}.

The suppression due to the EBL could be compensated by an additional emission component if the primary spectrum of injected gamma rays is hard enough.  Not including cascade effects, the photon index in the VHE range would need to be $\alpha \lesssim 2$.  However, for such a hard spectrum, the effects of cascades moving power from higher to lower energies cannot be neglected once there is enough luminosity at sufficiently high energies.  This may be realized by some population of blazars with a very hard spectrum.  Another possibility is that the cascades are initiated by VHE cosmic rays instead of gamma rays.  It is also possible that the ``VHE Excess" is not due to cascades, for example if the excess gamma rays are due to \textit{unaccounted-for} foreground emission within the Milky Way, where gamma-ray attenuation can be ignored.

VHDM annihilation or decay can lead to gamma-ray spectra that could explain the data~\cite[c.f.][]{aha+06}.  Extragalactic components would be cascaded, and could lead to similar final spectra as astrophysical sources, because cascade effects obscure the primary spectra.  Further, there could also be a Galactic component, which would not be attenuated or cascaded, and this can play an important role (see Section~4.1). 

Neutrinos can reach the Earth without attenuation.  Although the neutrino background has not been measured yet, neutrinos from dark matter annihilation or decay, as well as astrophysical neutrinos, may contribute to the background.  The resulting neutrino spectra are expected to be similar to those of the emission spectra that may extend to quite high energies, which can in principle reveal types of allowed channels.  As we stress below, taking into account both the gamma-ray and neutrino spectra is essential to probing VHDM models.

%%%%%%%%%%%%%%%%%%%%%%%%%%%%%%%%%%%%%%%%%%%%%%%%%%%%%%%%

\subsection{Calculation of cosmic backgrounds}

Now we describe how cosmic neutrino and gamma-ray backgrounds from VHDM are calculated.  Without attenuation or cascades, injected energies are higher than observed energies by $E'=(1+z)E$.
Here we focus on the extragalactic contribution, and below we note the importance of the foreground component from the Milky Way.  
For decaying dark matter, the received intensity (differential flux per energy, area, time, and solid angle) is 
\begin{equation}
\Phi = \frac{c}{4 \pi H_0} \int dz \,\,\, \frac{1}{\sqrt{\Omega_\Lambda+{(1+z)}^3 \Omega_m}} \,\,\, 
\frac{\rho_{\rm dm}}{m_{\rm dm} \tau_{\rm dm}} \frac{d S}{d E'},
\end{equation}
where $dS/dE'$ is the primary spectrum.  For annihilating dark matter, one has
\begin{equation}
\Phi = \frac{c}{4 \pi H_0} \int dz \,\,\, \frac{g(z) {(1+z)}^{3}}{\sqrt{\Omega_\Lambda+{(1+z)}^3 \Omega_m}} \,\,\, 
\frac{{<\sigma v> }_{\rm dm}\rho_{\rm dm}^2}{2 m_{\rm dm}^2} \frac{d S}{d E'}.
\end{equation}
For the decay case, the clustering of dark matter is irrelevant for the extragalactic contribution, whereas it is very important for the annihilation case.  The extragalactic clustering is represented by the flux-multiplier $g(z)$, which is~\cite{ts03}
\begin{equation}
g(z) = \int d M \,\,\, \frac{d n_{\rm halo}}{d M} g(c(M,z)) \frac{M}{\rho_{\rm dm}}  \frac{\Delta_c}{\Omega_{\rm dm}}.
\end{equation}
Here $d n_{\rm halo}/d M$ is the halo mass function, $c(M,z)$ is the concentration parameter and $\Delta_c$ is the overdensity relative to the critical density $\rho_c$.  With the Navarro-Frenk-White (NFW) profile~\cite{nfw95}, for the mass function obtained by Jenkins et al.~\cite{jen+01}, $g_0=g(z=0) \simeq {10}^4$ is obtained.  For the Press-Schechter mass function, $g_0 \simeq (4-5) \times {10}^4$~\cite{yuk+07}.  However, this factor may be largely enhanced by substructures.  For example, one of the most recent N-body simulations, Millenium II~\cite{bay+09}, leads to $g_0 \sim {10}^{6}-{10}^{7}$ when the luminosity function is extrapolated down to a damping scale mass limit of ${10}^{-6} h^{-1} M_{\odot}$.   This enhancement factor due to substructures can take a broad range of values~\cite[see also][]{zav+10,abd+10j,spr+08}.  We simply take $g_0={10}^6$ and we use the fitting function given by Yuksel et al.~\cite{yuk+07} for redshift evolution.  Other cases of substructure models can be easily compared because they scale the fluxes by a constant factor.

For the neutrino background, we numerically calculate it through Eqs.~(3.1) and (3.2) for $d S/dE'$ obtained using PYTHIA~\cite{sjo+06}.  For the extragalactic contribution to the DGB, we must calculate intergalactic cascades, and we perform detailed calculations by solving the Boltzmann equations taking into account pair creation, inverse Compton emission, synchrotron emission, and adiabatic loss (see Appendix A).  
Such detailed calculations are time-consuming but are important, especially in the high mass range, where VHE/UHE gamma rays are cascaded down as a result of many steps in the cascade process.
Then the extragalactic gamma-ray background flux is evaluated from
\begin{equation}
\Phi = \frac{c}{4 \pi H_0} \int \frac{dz}{(1+z) \sqrt{\Omega_\Lambda + (1+z)^3 \Omega_m}} \,\,\, \frac{d \dot{n}}{d E} (z),
\end{equation}
where the near-universal secondary spectrum observed at the Earth is used rather than the primary spectrum in the cosmic rest frame. 

Finally, the resulting background fluxes can be compared to the measured DGB or the sensitivity curves of neutrino detectors. 
In the next section, we show that the expected theoretical gamma-ray background can indeed be compatible to the VHE DGB measured by \textit{Fermi}.  In such scenarios, the corresponding neutrino background fluxes in some cases are high enough to detect with IceCube within a few years. 

%%%%%%%%%%%%%%%%%%%%%%%%%%%%%%%%%%%%%%%%%%%%%%%%%%%%%%%%

\subsection{Choices of final states}

The final states following dark matter annihilation or decay are unknown and model-dependent.  We try to develop general limits by considering some typical final states.  For the dark matter masses considered here, all standard model particles are kinematically available, and one generally expects that all possibilities will occur to some degree, though there may be some dominant channels.  As representative channels, we consider examples of gauge bosons, heavy quarks, and heavy leptons: ($b \bar{b}$, $W^+ W^-$, $\mu^+ \mu^-$), each with 100\% branching fractions.  (It is trivial to scale our results for any other assumed branching fractions.)

Once we show the results for these cases, we can discuss their similarities and differences, and say how to draw reasonably model-independent conclusions.  Of course, eventual more-detailed studies could further these conclusions.  The dependence of the spectra on mass exists but is not very strong, in the sense that constraints are sensitive to the typical energy flux at the typical energy.  In the neutrino case, the differences are visible but are moderate; in the gamma-ray case, the differences among the assumed final states can be much weaker due to cascades.  Thus, the choices made for the final states, while not easily justified in advance, are shown to be quite adequate.  For ease of comparison, we apply the same particle-antiparticle final states used for annihilation also to decay, as expected for scalar dark matter, even though these may not always be possible in other cases.  However, the results for a single one of those particles, or a similar particle in that class, would be very similar.

Some possible final states could seem quite different from the ones we assumed.  Either $\gamma \gamma$ or $e^+ e^-$ would yield no neutrino constraints, though they would have similar cascade bounds as each other.  Seemingly, $\nu \bar{\nu}$ would only lead to neutrino constraints but not gamma-ray constraints; however, electroweak bremsstrahlung leads to non-negligible electromagnetic branching ratios, and so gamma-ray constraints can be applied~\cite{anncong}.  (Note that those gamma-ray limits based on EGRET would be improved by $\sim 10$ by using {\it Fermi} data.)

To evaluate particle yields resulting from annihilation and decay, we run PYTHIA~\cite{sjo+06}.  At $\gtrsim$~PeV energy-scales, the yields at lower energies are scaled to provide a simple approximation to what might happen at higher energies.   While this is not perfectly correct due to the scaling violation or new physics beyond the standard model, our results show that this is sufficient for our purposes.

Examples of final state spectra are shown in Figures~3 and 4.  They are consistent with spectra shown in previous works~\cite[e.g.,][]{rot+11}, and satisfaction of energy conservation is checked.  Because only a fraction of the energy goes to baryons, the amount of electromagnetic and neutrino energy per annihilation or decay almost matches $m_{\rm dm} c^2$ in the decay case or $2 m_{\rm dm} c^2$ in the annihilation case.  For the $W^+ W^-$ channel, one sees a peak due to weak boson decay into a neutrino and lepton.  In addition, we consider the pure leptonic $\mu^+ \mu^-$ channel.  Such channels may be realized in some models such as a Sommerfeld-enhanced model including light force-scale carriers~\cite{ark+09}.  In Figure~3, we show the sum of contributions from gamma rays and electron-positron pairs, as appropriate for setting simple cascade bounds (see Section~5.2); in calculating secondary gamma-ray spectra below, we treat the primary gamma-ray and electron-positron spectra separately.

%++++++++++++++++++++++++++++++++++++++++++++++++++++++++++
\begin{figure*}[tb]
\begin{minipage}{0.49\linewidth}
\begin{center}
\includegraphics[width=\linewidth]{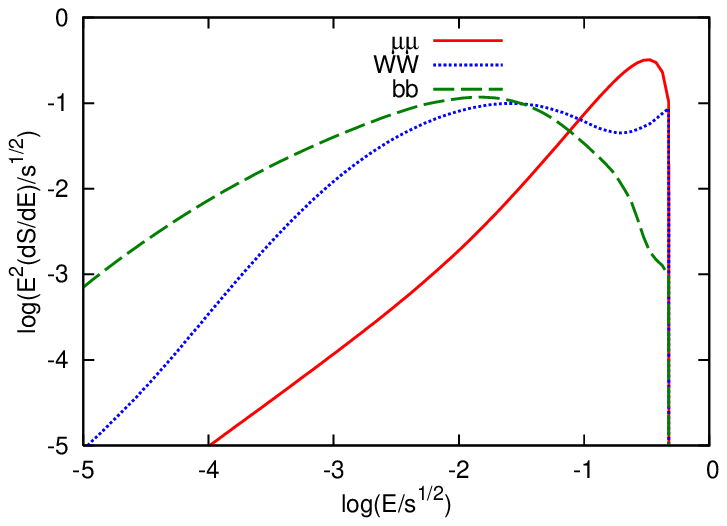}% Here is how to import EPS art
\caption{
Final state spectra of gamma rays/electrons/positrons (summed) for ${\rm DM} \rightarrow \mu^+ \mu^-$, ${\rm DM} \rightarrow W^+ W^-$, and ${\rm DM} \rightarrow b\bar{b}$ with $\sqrt{s}=100$~TeV.    
}
\end{center}
\end{minipage}
\begin{minipage}{.02\linewidth}
\end{minipage}
\begin{minipage}{0.49\linewidth}
\begin{center}
\includegraphics[width=\linewidth]{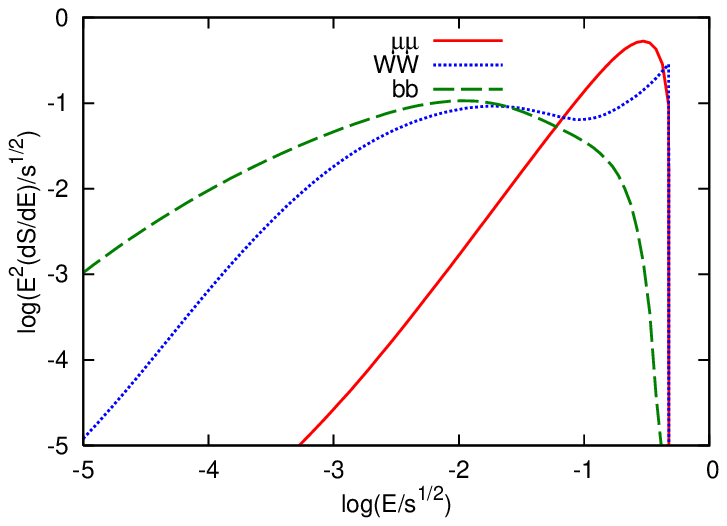}% Here is how to import EPS art
\caption{
Final state spectra of neutrinos (summed over flavors) for ${\rm DM} \rightarrow \mu^+ \mu^-$, ${\rm DM} \rightarrow W^+ W^-$, and ${\rm DM} \rightarrow b\bar{b}$ with $\sqrt{s}=100$~TeV.
\newline
}
\end{center}
\end{minipage}
\end{figure*}
%++++++++++++++++++++++++++++++++++++++++++++++++++++++++++

%%%%%%%%%%%%%%%%%%%%%%%%%%%%%%%%%%%%%%%%%%%%%%%%%%%%%%%%
%%%%%%%%%%%%%%%%%%%%%%%%%%%%%%%%%%%%%%%%%%%%%%%%%%%%%%%%

\section{VHDM signatures in the diffuse gamma-ray and neutrino backgrounds}

In this section, we show the results of gamma-ray and neutrino spectra calculated through the formulas provided in Section~3.  In particular, we compare the theoretical background flux to the measured DGB, and give some possible interpretations, first considering annihilation and then decay.

Throughout the paper, the low-energy part of the DGB is assumed to be due to astrophysical sources such as AGN and star-burst/star-forming galaxies.  For simplicity, we assume that it continues to higher energies and is attenuated (its cascade effects are negligible, because the spectrum is steep~\citep[e.g.,][]{ino11}), and we adopt the following parameterization, 
\begin{equation}
E_{\gamma}^2 \Phi_{\gamma} = 0.855 a \times {10}^{-7}~{\rm GeV}~{\rm cm}^{-2}~{\rm s}^{-1}~{\rm sr}^{-1}~{(E_{\gamma}/100~{\rm GeV})}^{2-\alpha} e^{-\tau_{\gamma \gamma}^{\rm eff}(E_\gamma, \alpha)},
\end{equation}
where $\tau_{\gamma}^{\rm eff} (E_\gamma, \alpha)$ is the effective intergalactic optical depth that is calculated via $\tau_{\gamma \gamma} (E_\gamma, z)$.
The real situation can be complicated by having several astrophysical contributions, especially from blazars and star-burst galaxies~\citep[see reviews][]{der12,dgb,ino11}.  Complete results would depend on many details, luminosity functions, spectral energy distributions, and so on. To avoid such complexities and uncertainties, the above form is assumed to be the sum of all astrophysical source types.  This is enough to demonstrate the potential importance of VHDM in the VHE DGB.  Detailed modeling and precise fitting procedures are beyond the scope of this work.

The discussions provided in this section should be regarded as the most optimistic case, in which the gamma-ray spectra from VHDM are a significant component of the observed DGB.  The constraints obtained in the next section are much more general and conservative, in that the allowed parameter space is defined by dark matter signals being less significant.

%%%%%%%%%%%%%%%%%%%%%%%%%%%%%%%%%%%%%%%%%%%%%%%%%%%%%%%%
 
\subsection{Possible dark matter scenarios}

We first consider annihilation of VHDM.  As an example, Figure~5 shows that the annihilation of $\sim 5$~TeV dark matter could seemingly explain the ``VHE Excess" in the DGB.  In this case, though the attenuation is moderately important, the cascade effect is not much relevant because secondary gamma rays appear at $\lesssim 10$~GeV.  The attenuation by the EBL is compensated by the fact that the initial spectrum is quite hard (c.f. Figure~3).  
Note that the implied cross section and the boost factor have to be quite large.  Even for $g_0={10}^7$, the corresponding cross section is ${<\sigma v>}_{\rm dm}= 9.5 \times {10}^{-25}~{\rm cm}^3~{\rm s}^{-1}$. 

As another example, Figure~6 shows a case where the intergalactic cascade is essential.  Information on the primary spectrum is smeared out, and the results do not depend on the dark matter mass as long as dark matter is heavy enough.  
Note that, in Figures~5 and 6, the dark matter contributions are basically subdominant and relevant only at $\gtrsim 20$~GeV.  Therefore, the required cross sections are slightly below current constraints on the basis of the \textit{Fermi} data on the DGB~\cite{abd+10j,aba+10}.   

%++++++++++++++++++++++++++++++++++++++++++++++++++++++++++
\begin{figure*}[bt]
\begin{minipage}{0.49\linewidth}
\begin{center}
\includegraphics[width=\linewidth]{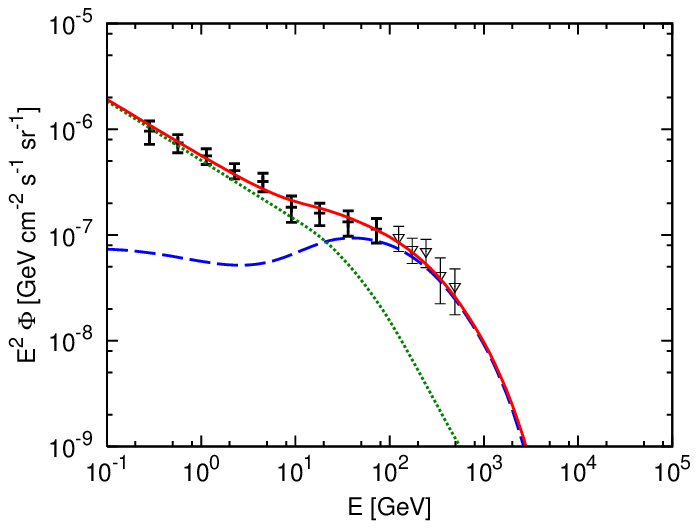}% Here is how to import EPS art
\caption{
The isotropic DGB (sum; solid curve) from dark matter (dashed curve) and an EBL-attenuated power-law extragalactic component (dotted curve).  The assumed dark matter energy budget is $Q_{\rm dm}=3.0 \times {10}^{45}~{\rm erg}~{\rm Mpc}^{-3}~{\rm yr}^{-1}$ for the $W^+ W^-$ channel.  For annihilating dark matter, the corresponding dark matter mass is $m_{\rm dm} c^2=5$~TeV.  For the EBL-attenuated power-law component, $a=0.9$ and $\alpha=2.56$ are assumed with star formation evolution.    
}
\end{center}
\end{minipage}
\begin{minipage}{.02\linewidth}
\end{minipage}
\begin{minipage}{0.49\linewidth}
\begin{center}
\includegraphics[width=\linewidth]{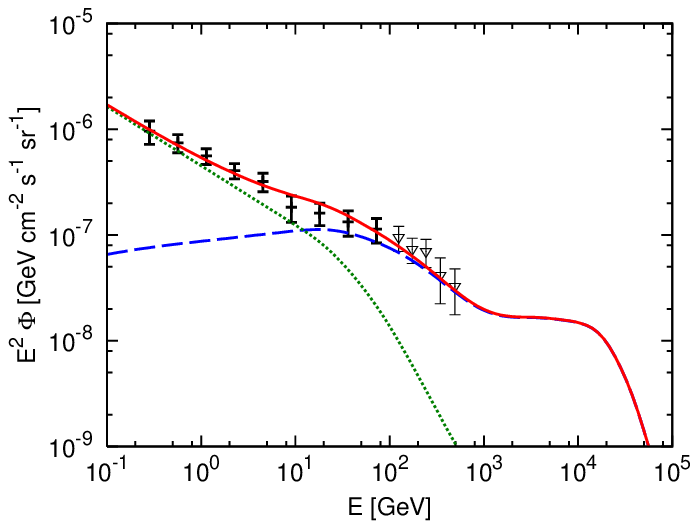}% Here is how to import EPS art
\caption{The same as Figure~5, but $Q_{\rm dm}=3.8 \times {10}^{45}~{\rm erg}~{\rm Mpc}^{-3}~{\rm yr}^{-1}$ is used for the $b \bar{b}$ channel.  For annihilating dark matter, the corresponding dark matter mass is $m_{\rm dm} c^2=5$~EeV, but resulting secondary spectra do not depend on the mass or on detailed primary spectra due to the intergalactic cascade.  For the EBL-attenuated power-law component, $a=0.8$ and $\alpha=2.56$ are assumed.
\newline}
\end{center}
\end{minipage}
\end{figure*}
%++++++++++++++++++++++++++++++++++++++++++++++++++++++++++

In both the cases, quite large values of the boost factor and/or the annihilation cross section are required in order that VHDM annihilation makes a significant contribution to the DGB. 
The annihilation cross section could be larger than canonical values in some models, e.g., the Sommerfeld-enhanced model~\cite{ark+09}, when dark matter is nonthermally produced.  
Large boost factors also seem necessary, which can be realized by significant substructure contributions.  The substructure can enhance both the Galactic and extragalactic contributions, and the extragalactic contribution can be more important in some optimistic substructure models~\citep[see][and references therein]{abd+10j}.  For demonstrative purposes, we consider only the extragalactic contribution, which also gives more conservative constraints compared to the case where both the contributions are included.  However, since the large cross section and/or large boost factor are required, they may conflict with other constraints~\cite{abr+11,dwarfann,mb12}. 

Because of the above caveats, in this paper, we mainly focus on decaying VHDM.  
For decaying dark matter, the flux is insensitive to substructures, and both the Galactic and extragalactic contributions should be relevant.  In Appendix B, we describe how the Galactic contributions are calculated.  Introducing the $\mathcal J$ factor, i.e., the intensity of the Galactic contribution that is proportional to the line of sight integration of the dark matter density, we can define ${\mathcal J}_{\Omega}$ as the quantity averaged over the sky region~\cite{yuk+07}, which is used for the estimate of the Galactic contribution.  

In Figure~7, we show an example of a DGB produced by decaying VHDM with $m_{\rm dm} c^2 \simeq 3$~TeV for the ${\rm DM} \rightarrow W^+ W^-$ channel.  
It suggests that the sum of the two components can explain the VHE DGB if the dark matter mass is in the appropriate range and the extragalactic astrophysical component is expressed by a simple power law~\footnote{It is also consistent with present constraints from anti-proton measurements~\cite{cho11}.}.  Though the Galactic component is naively expected to be comparable to the extragalactic one, it is actually dominant here.  As indicated in Figure~7, the extragalactic component is suppressed by interactions with the EBL at $\gtrsim 100$~GeV energies.  On the other hand, secondary gamma rays by primary electrons/positrons and absorbed $\lesssim$~TeV gamma rays are radiated in the GeV range, where the astrophysical power-law component is supposed to be dominant.

We show the cases with ${\rm DM} \rightarrow \mu^+ \mu^-$ in Figures~8 and 9, where the Galactic component is dominant in the relevant energy range.  In Figure~8, the hard spectral component in the $\sim 100$~GeV range is dominated by secondary inverse Compton emission rather than primary gamma-ray emission from decaying dark matter.  In Figure~9, the Galactic synchrotron component is dominant since UHE pairs mainly cool via synchrotron emission rather than inverse Compton emission.  The latter situation is possible only when VHDM is super heavy.  In both the cases, the extragalactic component also has a similar spectrum, though it has a tail due to the intergalactic cascade.

%++++++++++++++++++++++++++++++++++++++++++++++++++++++++++
\begin{figure*}[bt]
\begin{minipage}{0.49\linewidth}
\begin{center}
\includegraphics[width=\linewidth]{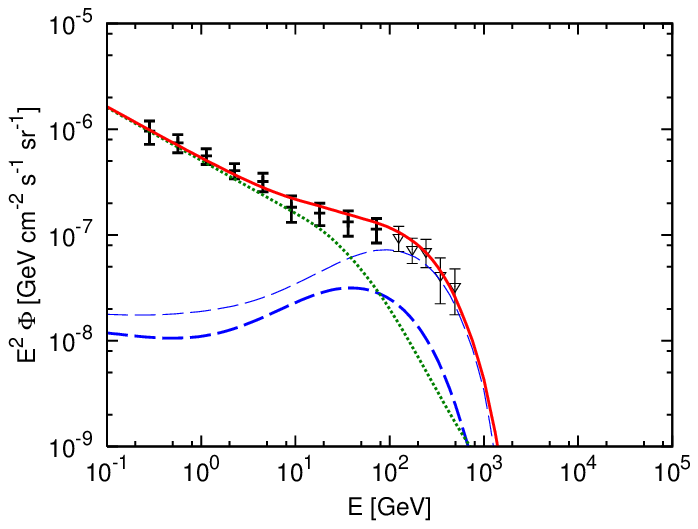}% Here is how to import EPS art
\caption{
The DGB (sum; solid curve) from decaying dark matter (dashed curves) and an EBL-attenuated power-law extragalactic component (dotted curve).  We use $m_{\rm dm} c^2 \simeq 3$~TeV and $\tau_{\rm dm}=1.2 \times {10}^{27}$~s for ${\rm DM} \rightarrow W^+ W^-$.  Both the extragalactic (thick curve) and galactic components (thin curve) are shown.  For the power-law, $a=0.9$ and $\alpha=2.50$ are assumed with star formation evolution.      
}
\end{center}
\end{minipage}
\begin{minipage}{.02\linewidth}
\end{minipage}
\begin{minipage}{0.49\linewidth}
\begin{center}
\includegraphics[width=\linewidth]{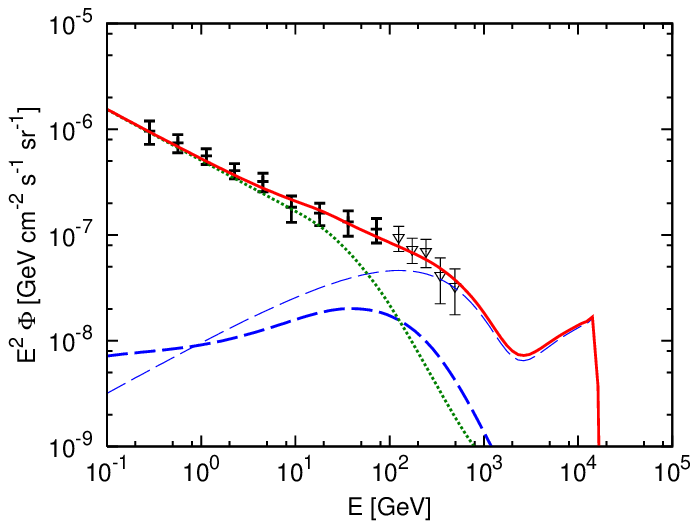}% Here is how to import EPS art
\caption{The same as Figure~7, but for $m_{\rm dm} c^2 \simeq 30$~TeV and $\tau_{\rm dm}=2.0 \times {10}^{27}$~s, and ${\rm DM} \rightarrow \mu^+ \mu^-$.  For the power-law component, $a=0.9$ and $\alpha=2.48$ are assumed.
\newline \, \newline \, \newline \, \newline \, \newline
}
\end{center}
\end{minipage}
\end{figure*}
%++++++++++++++++++++++++++++++++++++++++++++++++++++++++++

%++++++++++++++++++++++++++++++++++++++++++++++++++++++++++
\begin{figure*}[bt]
\begin{minipage}{0.49\linewidth}
\begin{center}
\includegraphics[width=\linewidth]{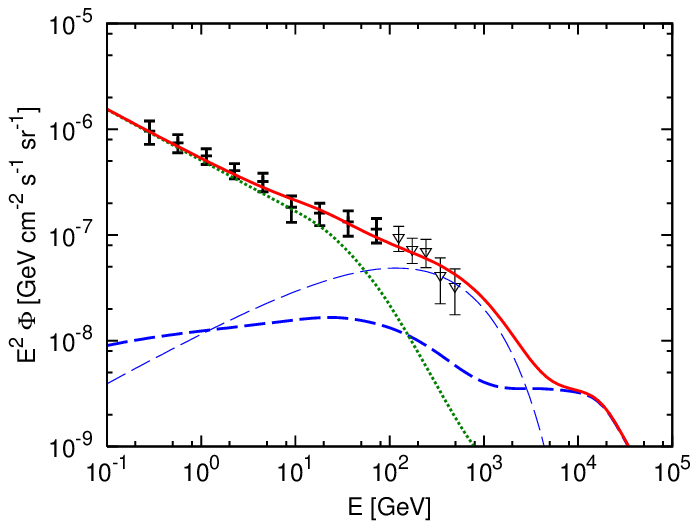}% Here is how to import EPS art
\caption{The same as Figure~7, but for $m_{\rm dm} c^2={10}$~EeV and $\tau_{\rm dm}=1.8 \times {10}^{27}$~s, and ${\rm DM} \rightarrow \mu^+ \mu^-$.  For the power-law component, $a=0.9$ and $\alpha=2.48$ are assumed. 
}
\end{center}
\end{minipage}
\begin{minipage}{.02\linewidth}
\end{minipage}
\begin{minipage}{0.49\linewidth}
\begin{center}
\includegraphics[width=\linewidth]{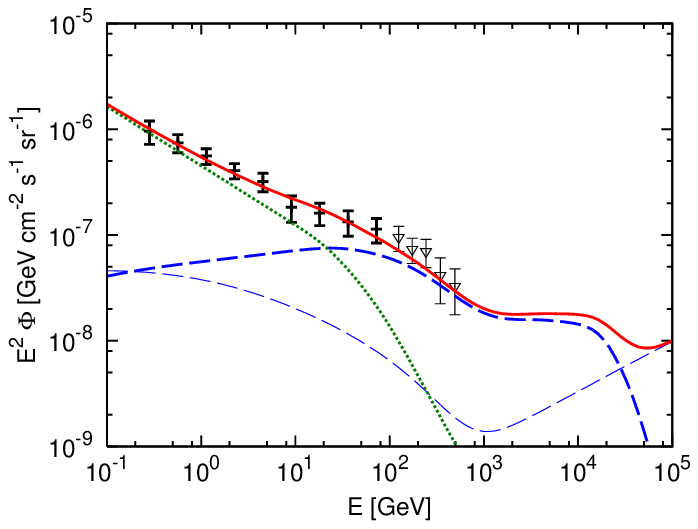}% Here is how to import EPS art
\caption{The same as Figure~7, but for $m_{\rm dm} c^2=10$~EeV, $\tau_{\rm dm}=4.7 \times {10}^{26}$~s and ${\rm DM} \rightarrow b \bar{b}$.  For the power-law component, $a=0.8$ and $\alpha=2.56$ are assumed.}
\end{center}
\end{minipage}
\end{figure*}
%++++++++++++++++++++++++++++++++++++++++++++++++++++++++++

%++++++++++++++++++++++++++++++++++++++++++++++++++++++++++
\begin{figure*}[bt]
\begin{minipage}{0.49\linewidth}
\begin{center}
\includegraphics[width=\linewidth]{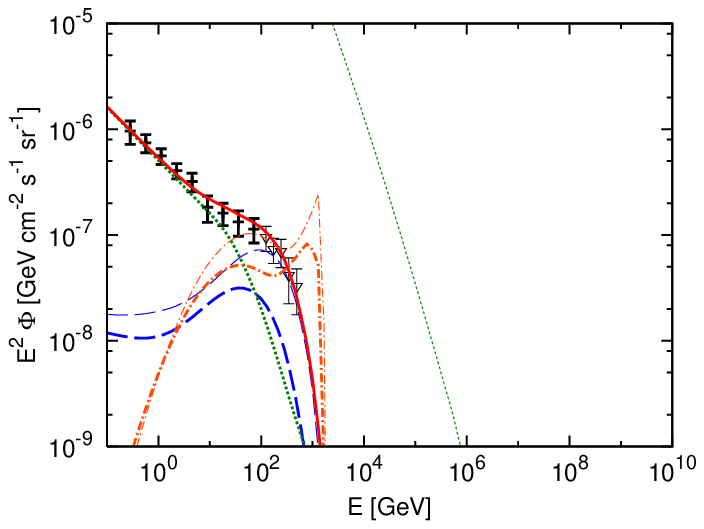}% Here is how to import EPS art
\caption{The isotropic DGBs (dashed curves) and neutrino backgrounds (dot-dashed curves) of decaying dark matter.  The dark matter parameters are the same as those in Figure~7.  Both the extragalactic (thick curves) and galactic components (thin curves) are shown.  The atmospheric neutrino background is also shown (dotted curve).  This case is allowed.
}
\end{center}
\end{minipage}
\begin{minipage}{.02\linewidth}
\end{minipage}
\begin{minipage}{0.49\linewidth}
\begin{center}
\includegraphics[width=\linewidth]{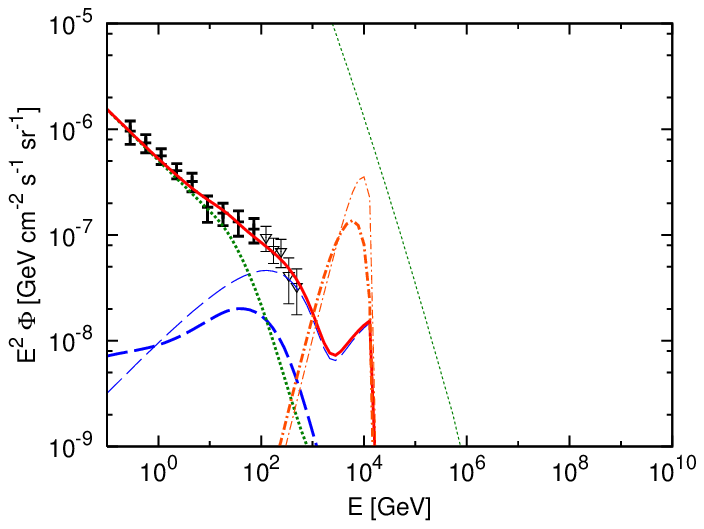}% Here is how to import EPS art
\caption{The same as Figure~11, but the dark matter parameters come from Figure~8.  This case is allowed now, but is testable by the full IceCube. 
\newline \, \newline \,  \newline \, \newline}
\end{center}
\end{minipage}
\end{figure*}
%++++++++++++++++++++++++++++++++++++++++++++++++++++++++++
%++++++++++++++++++++++++++++++++++++++++++++++++++++++++++
\begin{figure*}[bt]
\begin{minipage}{0.49\linewidth}
\begin{center}
\includegraphics[width=\linewidth]{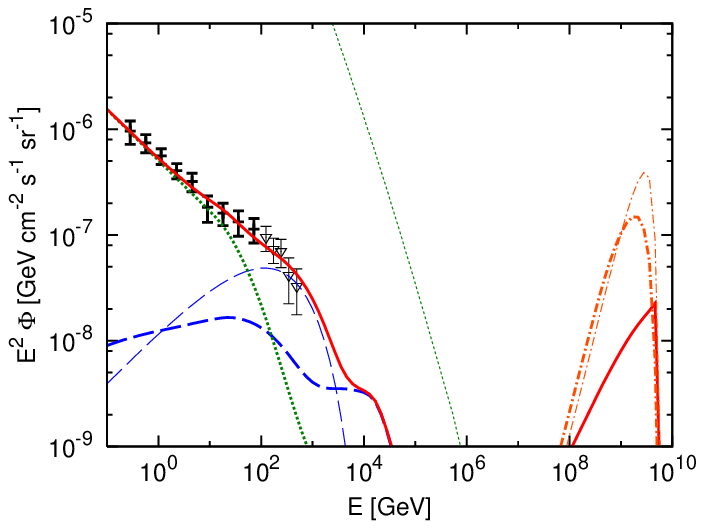}% Here is how to import EPS art
\caption{The same as Figure~11, but the dark matter parameters come from Figure~9.  This case is ruled out by neutrino observations. 
\newline
}
\end{center}
\end{minipage}
\begin{minipage}{.02\linewidth}
\end{minipage}
\begin{minipage}{0.49\linewidth}
\begin{center}
\includegraphics[width=\linewidth]{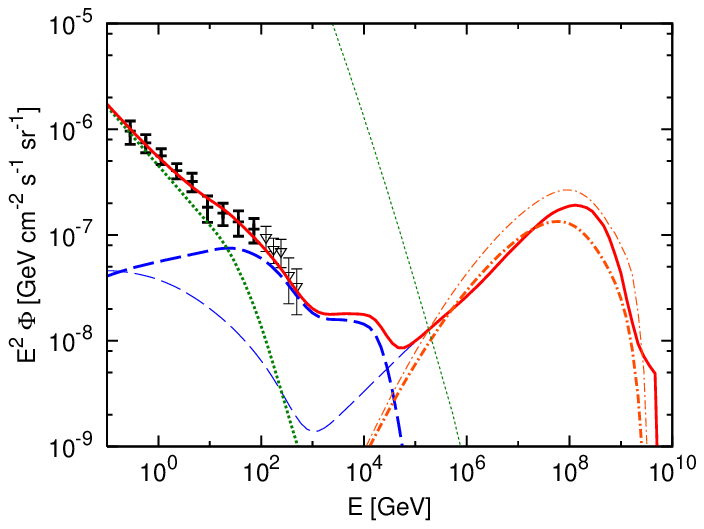}% Here is how to import EPS art
\caption{The same as Figure~11, but the dark matter parameters come from Figure~10.  This case is ruled out by neutrino observations. 
\newline}
\end{center}
\end{minipage}
\end{figure*}
%++++++++++++++++++++++++++++++++++++++++++++++++++++++++++

Alternatively, as in Figure~10, one may consider cases where the extragalactic contribution is more relevant.  When the Galactic component is present, this is possible only when dark matter is so heavy that the Galactic component exceeds the DGB measured by \textit{Fermi}.  In this case, as in Figure~6, the gamma-ray emission at $\sim 100$~GeV energies is dominated by the VHDM-induced intergalactic cascade.  Since the cascade spectrum is not much sensitive to the primary spectrum, this intergalactic cascade scenario is allowed only in exotic cases where dark matter mass is higher than $\sim 10$~PeV.  Well-motivated particle models typically lie below $\sim 100$~TeV, but super heavy dark matter is still viable and the intergalactic cascade spectrum is similar even for masses $\gtrsim {10}^{9}$~GeV.

%%%%%%%%%%%%%%%%%%%%%%%%%%%%%%%%%%%%%%%%%%%%%%%%%%%%%%%%

\subsection{Importance of neutrino observations}

In the previous subsection, we discussed possible cases where VHDM contributes significantly to the VHE DGB.  It would be interesting to consider how we can test those scenarios.  As discussed in Section~2, for channels involving quarks and massive bosons, pions and other mesons are produced as a result of hadronization, and these lead to neutrinos.  For heavy leptonic channels, neutrinos are produced via decay of muons and taus.  Those neutrinos should not violate neutrino constraints as discussed in Section~3. 

In Figures~11-14, we show neutrino spectra in addition to the gamma-ray spectra (note the larger energy range).  In the Galactic scenario demonstrated in Figures~11 and 12, the required dark matter mass is in the range of $\sim 3-30$~TeV.  Then the corresponding neutrino background is below the atmospheric neutrino flux, and the neutrino limits do not apply.  In the case of ${\rm DM} \rightarrow W^+ W^-$, neutrinos would not be seen by IceCube (see Section~5.2).  The case of ${\rm DM} \rightarrow \mu^+ \mu^-$ may be more interesting.  While it is still consistent with recent limits~\cite{rot+11}, the signal can be seen by the full IceCube.  The non-detection can rule out this possibility within three years, as indicated in Section~5.2.        

As suggested in Figures~13 and 14, neutrino observations are more powerful when the dark matter mass is heavier than $\sim 10$~PeV.  Such cases include the VHDM-induced cascade scenario demonstrated in Figures~14, where the intergalactic cascade component is dominant.  
Unless we consider channels without neutrinos (e.g., ${\rm DM} \rightarrow e^+ e^-$), however, some of these exotic scenarios already violate the neutrino constraints.  Indeed, the constraints provided in Section~5.1 suggests that, for such heavy dark matter masses, the neutrino constraints are more stringent than the cascade gamma-ray constraints.

%%%%%%%%%%%%%%%%%%%%%%%%%%%%%%%%%%%%%%%%%%%%%%%%%%%%%%%%
%%%%%%%%%%%%%%%%%%%%%%%%%%%%%%%%%%%%%%%%%%%%%%%%%%%%%%%%

\section{General gamma-ray and neutrino constraints on VHDM}

In this section, we show our general constraints on dark matter properties.  These results are more conservative; the cases presented in the previous section are on the boundary of those being excluded.

Cascade gamma-ray constraints are placed by requiring the calculated cascade gamma-ray spectrum not exceed the measured DGB data at any individual energy bin by more than given significance~\cite{abd+10j}.  When dark matter mass lies in $\lesssim$~TeV energy-scales, we find that the obtained constraints are consistent with the previous work~\citep[e.g.,][]{aba+10}.  We extend the constraints to higher masses, by taking account of cosmic cascades.  

The gamma-ray constraints will be improved if more point sources contributing to the DGB are resolved, but the expected improvements are moderate~\cite{aba+10}.  On the other hand, neutrino constraints can be improved with time.  Hereafter we focus on ``forecasted" neutrino constraints to demonstrate the power of neutrino observations, assuming non-detections of neutrinos with full IceCube three-year observations.  They are not really limits yet, and the neutrino background may be seen in the near future~\cite{neu,mur11}. 

How the neutrino constraints are placed depends on energy.  
We estimate the muon event rate by
\begin{equation}
{\mathcal N}_{\mu} (\geq E_\nu) = \Omega T  \int dE_{\nu}^\prime \,\,\, A_{\rm eff} (E_{\nu}^\prime) \Phi_{\nu} (E_\nu^\prime), 
\end{equation}
where $\Omega T$ is the exposure and $A_{\rm eff} (E_\nu)$ is the neutrino effective area.  For the dark matter signals, the neutrino mixing is taken into account assuming $\theta_{12}=0.59$ and $\theta_{23}=\pi/4$~\cite{sv06}.  For the background events, we assume the conventional atmospheric muon neutrino background to calculate $N_{\rm bkg}$~\cite{neu}, with the assumption that any non-terrestrial neutrino background is not observed.  The full IceCube effective area is assumed to be three times as that of IceCube-40~\cite{ahr+04}, and constraints are set by the criterion, $N_{\rm sig}/\sqrt{N_{\rm sig}+N_{\rm bkg}} < \delta$, where $\delta$ is the Gaussian significance.  

At low energies (at least in the $\lesssim 300$~TeV range), the atmospheric neutrino flux is dominant, and the constraints will improve as the square root of time.  At higher energies, the atmospheric neutrino flux becomes negligible, and the constraints will improve linearly with time unless some other background is detected.  

At further high energies, the Earth becomes optically thick for neutrinos, detecting down-going and horizontal neutrinos is important.  The above criterion would be rather conservative.  In order to have better constraints at sufficiently high masses, we adopt the quasi-differential sensitivity obtained by the analysis performed at extremely high energies~\cite{abb+11b}.

%%%%%%%%%%%%%%%%%%%%%%%%%%%%%%%%%%%%%%%%%%%%%%%%%%%%%%%%

\subsection{Annihilation}

%++++++++++++++++++++++++++++++++++++++++++++++++++++++++++
\begin{figure}[tb]
\begin{center}
\includegraphics[width=0.7\textwidth]{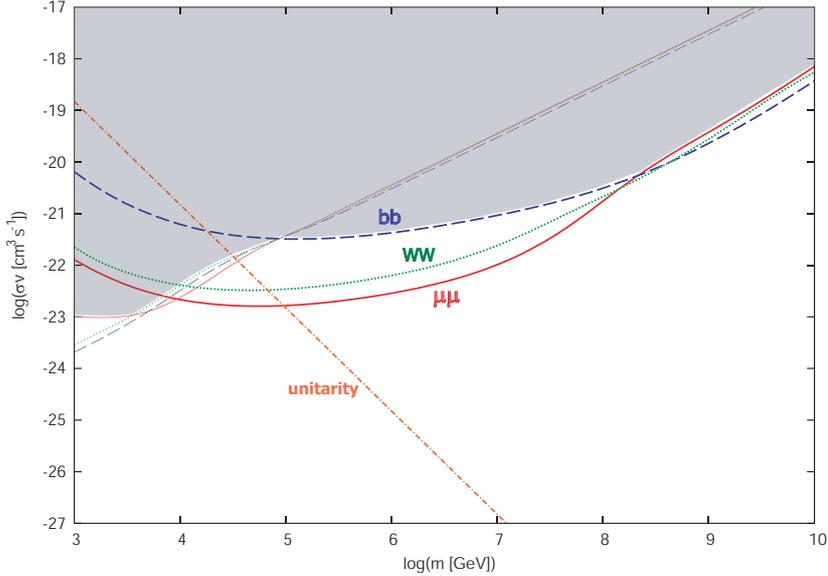}
\end{center}
\caption{Neutrino (thick curves) and cascade gamma-ray (thin curves) constraints on the annihilation cross section of VHDM.  Non-detections in IceCube three-year observations and \textit{Fermi} DGB measurements are used, respectively. Only the extragalactic contribution is taken into account, and constraints would be weaker for more pessimistic substructure models.  The shaded region is defined by the more stringent bound between the neutrino and gamma-ray bounds, where for each we choose the most conservative limit among the selected limits for the three channels. 
}
\end{figure}
%++++++++++++++++++++++++++++++++++++++++++++++++++++++++++

Our results for annihilating VHDM are shown in Figure~15.  
As in the previous section, we have assumed the large boost due to substructures, which could allow the extragalactic contribution to dominate over the Galactic contribution~\citep[see][and references therein]{aba+10}.  Then, the gamma-ray constraints below $\lesssim 10$~TeV are consistent with previous works~\cite[e.g.,][]{aba+10}.  But, due to detailed calculations of intergalactic cascades, our results extend to much higher masses.  
Without this assumption, other constraints can give stronger results, though they are not focus of this work.  For example, constraints from dwarf galaxies give $\lesssim {10}^{-24}-{10}^{-23}~{\rm cm}^{3}~{\rm s}^{-1}$ around 1~TeV, depending on channels~\cite{dwarfann}.  The constraint is weaker as $\propto m_{\rm dm}$ as long as the energy flux limit is similar.  Other limits including IACT constraints also give interesting limits~\citep[see a review][and references therein]{por+11}, and the recent HESS measurement in particular provided $\lesssim 3 \times {10}^{-25}~{\rm cm}^{3}~{\rm s}^{-1}$ at TeV for the quark-antiquark channel.  Note that the isotropic DGB is difficult for IACTs to measure.  
For the neutrino constraints, previous limits were obtained for cases where the substructure contribution is moderate and the Galactic contribution is dominant.  In such cases, the optimized searches for the Galactic halo give better constraints, and IceCube-22 results gave $\lesssim {10}^{-22}~{\rm cm}^{3}~{\rm s}^{-1}$ for the $\mu^+ \mu^-$ channel and $\lesssim {10}^{-20}~{\rm cm}^{3}~{\rm s}^{-1}$ for the $b \bar{b}$ channel~\cite{rot+11}.  Although our limits shown in Figure~15 are apparently better, they are more model-dependent. 

For the annihilation of point-like particles, whether dark matter is produced thermally or nonthermally, the unitarity bound~\citep[see,e.g.,][]{annuni} can be applied.  When the annihilation cross section is $s$-wave dominated, one has~\cite{bea+07}
\begin{equation}
{<\sigma v>}_{\rm dm} \leq \frac{4 \pi \hbar^2}{m_{\rm dm}^2 v} \simeq 1.5 \times {10}^{-19}~{\rm cm}^{3}~{\rm s}^{-1}~{\left( \frac{m_{\rm dm} c^2}{\rm TeV}\right)}^{-2} {\left( \frac{v}{300~{\rm km}~{\rm s}^{-1}} \right)}^{-1},
\end{equation}
which may restrict the annihilating cross section of super heavy dark matter.  We also show this unitarity bound in Figure~15.  The unitarity bound can be avoided if the dark matter is not point-like.  Note that dark matter with high cross sections in the present-day universe may have been produced nonthermally in earlier times.

Beacom et al.~\cite{bea+07} argued that neutrino limits give the most conservative bounds on the total annihilation cross section, because the neutrino limits are typically weaker than those with gamma rays.  In the typically-considered mass range, that principle holds in this example.  Interestingly, when a larger range of masses is considered, that can change.  The present results show that the neutrino limits are stronger than the gamma-ray limits at very high masses.  The reason is that the neutrino experiments are directly sensitive to VHE neutrinos, while sufficiently high-energy gamma rays are cascaded down to lower energies, where there are more gamma rays from other causes.  Therefore, for sufficiently high dark matter masses, the logic about the most conservative way to set a limit on the total cross section would change, and the cascade gamma-ray bound could be used to put constraints on the total cross section.

The shaded region is depicted as follows.  In comparing these gamma and neutrino limits, we take the stronger of the two.  Then, we conservatively take the worst limit among various channels.  Note that the differences between the gamma-ray limits for the different channels are typically small, whereas this matters more in the neutrino limits.  We are conservative about the channels that are uncertain, whereas we may use the stronger limit because both the messengers are reliable.

For annihilating VHDM, we see that analytical estimates basically agree with the numerical calculations. 
Using $Q_\gamma \sim f_{\rm em} Q_{\rm dm}$ (where $f_{\rm em}$ is the bolometric fraction in the electromagnetic component), one has (for gamma rays)
\begin{eqnarray}
{<\sigma v>}_{\rm dm} &\lesssim& 2.9 \times {10}^{-20}~{\rm cm}^{3}~{\rm s}^{-1}~\left(\frac{m_{\rm dm} c^2}{10~{\rm PeV}} \right) \nonumber \\
&\times& {\left( \frac{0.113}{\Omega_{\rm dm} h^2}\right)}^2 {\left( \frac{1.05 \times {10}^{5}~{\rm GeV}~{\rm cm}^{-3}}{\rho_c c^2 h^{-2}} \right)}^2 \left( \frac{{10}^6}{g_0} \right) \left( \frac{f_{\rm em}^{-1}}{3} \right) \xi_z^{-1}.
\end{eqnarray}
Also, using $E_{\nu} Q_{E_\nu} \sim Q_{\rm dm}/{\mathcal R}_\nu$ (where $E_\nu^2 (d S_\nu/dE_\nu) \equiv \sqrt{s}/{\mathcal R}_{\nu}$), one has (for neutrinos)
\begin{eqnarray}
{<\sigma v>}_{\rm dm} &\lesssim& 2.1 \times {10}^{-22}~{\rm cm}^{3}~{\rm s}^{-1}~\left(\frac{E_\nu^2 \Phi_\nu^{\rm lim}}{3 \times {10}^{-9}~{\rm GeV}~{\rm cm}^{-2}~{\rm s}^{-1}~{\rm sr}^{-1}} \right)
\left(\frac{m_{\rm dm} c^2}{10~{\rm PeV}} \right)  \nonumber \\ 
&\times& {\left( \frac{0.113}{\Omega_{\rm dm} h^2}\right)}^2 {\left( \frac{1.05 \times {10}^{5}~{\rm GeV}~{\rm cm}^{-3}}{\rho_c c^2 h^{-2}} \right)}^2 \left( \frac{{10}^6}{g_0} \right) ({\mathcal R}_{\nu}/10) \xi_z^{-1}.
\end{eqnarray}
Note that, as seen in Figure~15, the constraints become weaker as dark matter is heavier.  

Generally speaking, the ratio of the Galactic contribution to the extragalactic contribution is expressed as~\cite{yuk+07}
\begin{equation}
\frac{\Phi_{\rm G}}{ \Phi_{\rm EG}} \approx \frac{B {\mathcal J}_{\Omega} R_{\rm sc} \rho_{\rm sc}^2}{c t_H  \rho_{\rm dm}^2 g_0 \xi_z} \sim 0.1 \frac{B {\mathcal J}_{\Omega}}{(g_0/{10}^6) \xi_z},
\end{equation}
where $B$ is the boost factor for the Galactic contribution.  Note that, for gamma rays, the denominator can be affected by propagation effects.  While the boost toward the Galactic center is order of unity, the overall Galactic boost factor can take a wide range of values~\citep[see, e.g.,][]{zav+10}, so that the above ratio highly depends on substructure models.  
In the previous section, we discuss possibilities that VHDM contributes to the DGB, where significant boosts due to substructures are assumed for annihilating dark matter.  To avoid complexities due to Galactic substructures, for obtaining the constraints, we do not include the Galactic contribution in the present study, which means Figure~15 gives conservative constraints for a given $g_0$.  
In the absence of substructures, the extragalactic contribution is typically subdominant~\cite{yuk+07}, and the constraints from regions around the Galactic center should be more stringent~\cite{anngs}.  

%%%%%%%%%%%%%%%%%%%%%%%%%%%%%%%%%%%%%%%%%%%%%%%%%%%%%%%%

\subsection{Decay}

In Figure~16, we show our numerical results for present-day decaying VHDM.
The gamma-ray constraints give $\tau_{\rm dm} \gtrsim {10}^{26}-{10}^{27}$~s.  For the ${\rm DM} \rightarrow \mu^+ \mu^-$ channel, our gamma-ray constraint is consistent with the previous result that is obtained only at $< 10$~TeV~\citep[e.g.,][]{aba+10,deccong2}.  
Our result is applicable even at higher masses, thanks to detailed calculations of intergalactic cascades.  One sees that the cascade gamma-ray bound is largely mass-independent at sufficiently high masses.  This is because the energy injection rate of decaying dark matter does not depend on $m_{\rm dm}$ (because the number density scales as $\propto 1/m_{\rm dm}$), and the cascade spectrum is near-universal so the DGB constrains the bolometric electromagnetic energy budget without depending on details of final state spectra.  
Also, the cascade bound is reasonably channel-independent.  While this is obvious when the energy fraction into electromagnetic components of decay products is similar, it may not be true e.g., for the $\nu \bar{\nu}$ channel even if all decay products are standard model particles.  However, even in such cases, the cascade gamma-ray bound can be effectively applied when it can be regarded as a more conservative constraint compared to the neutrino constraint.  

The gamma-ray constraints are more stringent than neutrino ones at lower masses, whereas the neutrino constraints become stronger for masses of $\gtrsim 10-100$~TeV (as above for Ref.~\cite{bea+07} and annihilation, this reverses the logic for Ref.~\cite{pal08} and decay).  In other words, neutrino observations give conservative limits at $\lesssim 15$~TeV for ${\rm DM} \rightarrow \mu^+ \mu^-$, $\lesssim 25$~TeV for ${\rm DM} \rightarrow W^+ W^-$, $\lesssim 100$~TeV for ${\rm DM} \rightarrow b \bar{b}$, respectively. 
Compared to the gamma-ray constraints, the neutrino constraints are sensitive to the choice of decay channels.  This is because the differential spectrum is relevant for neutrino observations, and harder neutrino emission is more favored to avoid the atmospheric neutrino background and to be detected with a larger effective area.  
Our results are basically consistent with previous limits that are obtained at $<100$~TeV where the atmospheric neutrino background is measured~\cite{pal08,decconn}, but we also cover the higher mass range.  And our constraints are expected to be slightly better than those of Ref.~\cite{pal08} because the ratio of the signal to the background increases with time. 
Also, the forecasted limits are expected to be much better than the previous IceCube-22 results unless the diffuse neutrino background is discovered~\cite{rot+11}.  
Note that, of course, the lifetime for the cases shown in Figures~5-14 is consistent with the limits shown in Figure~16. 

%++++++++++++++++++++++++++++++++++++++++++++++++++++++++++
\begin{figure}[tb]
\begin{center}
\includegraphics[width=0.7\textwidth]{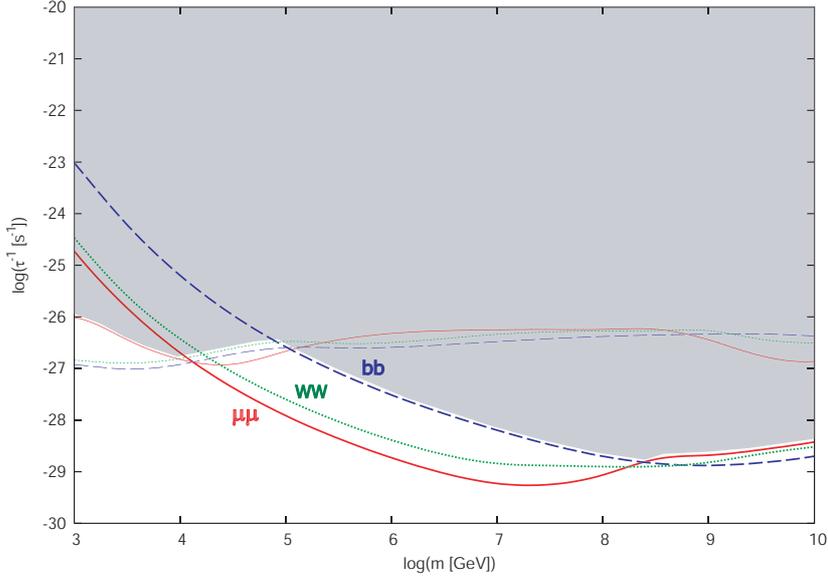}
\end{center}
\caption{Neutrino (thick curves) and cascade gamma-ray (thin curves) constraints on the lifetime of decaying VHDM (90~\% CL).  Non-detections in IceCube three-year observations and \textit{Fermi} DGB measurements are used, respectively.  Both of the Galactic and extragalactic contributions are included. 
The shaded region is defined by the more stringent bound between the neutrino and gamma-ray bounds, where for each we choose the most conservative limit among the selected limits for the three channels. 
}
\end{figure}
%++++++++++++++++++++++++++++++++++++++++++++++++++++++++++

For decaying dark matter at higher masses, the cascade gamma-ray bound becomes more conservative than the neutrino limits.  As discussed in the previous section, the cascade leads to a near-universal spectrum, so the DGB constraints are sensitive to the bolometric energy budget rather than the differential energy budget, and details of the primary spectra do not matter for the constraints.  Therefore, as in annihilating dark matter, the cascade gamma-ray bound could be used to put constraints on the total decay rate at sufficiently high masses.   

Our numerical results roughly agree with analytical expectations for the cosmic limits.  Using $Q_\gamma \sim f_{\rm em} Q_{\rm dm}$ (where $f_{\rm em}$ is the bolometric fraction in the electromagnetic component), we have (for gamma rays)
\begin{equation}
\tau_{\rm dm}^{-1} \lesssim 3.4 \times {10}^{-27}~{\rm s}^{-1}~\left( \frac{0.113}{\Omega_{\rm dm} h^2}\right) \left( \frac{1.05 \times {10}^{5}~{\rm GeV}~{\rm cm}^{-3}}{\rho_c c^2 h^{-2}} \right) 
\left( \frac{f_{\rm em}^{-1}}{3} \right) \xi_z^{-1},
\end{equation}
where the constraints are expected to be largely mass-independent for present-day decaying VHDM~\citep[e.g.,][]{gon92}.  
Also, using $E_{\nu} Q_{E_\nu} \sim Q_{\rm dm}/{\mathcal R}_\nu$ (where $E_\nu^2 (d S_\nu/dE_\nu) \equiv \sqrt{s}/{\mathcal R}_{\nu}$), one has (for neutrinos)
\begin{eqnarray}
\tau_{\rm dm}^{-1} &\lesssim& 2.4 \times {10}^{-29}~{\rm s}^{-1}~\left(\frac{E_\nu^2 \Phi_\nu^{\rm lim}}{3 \times {10}^{-9}~{\rm GeV}~{\rm cm}^{-2}~{\rm s}^{-1}~{\rm sr}^{-1}} \right) \nonumber \\ 
&\times& \left( \frac{0.113}{\Omega_{\rm dm} h^2}\right) \left( \frac{1.05 \times {10}^{5}~{\rm GeV}~{\rm cm}^{-3}}{\rho_c c^2 h^{-2}} \right) 
({\mathcal R}_{\nu}/10) \xi_z^{-1}.
\end{eqnarray}

The cascade gamma-ray bound is simply understood through Eq.~(5.6).  
For demonstration, based on the result in Figure~1, we also apply the energy budget constraint in the no redshift evolution case, $Q_\gamma \approx 2 .6\times {10}^{45}~{\rm erg}~{\rm Mpc}^{-3}~{\rm yr}^{-1}$, to decaying VHDM.  
Results are shown in Figure~17.  One sees that the simple cascade gamma-ray constraints are comparable to the detailed constraints shown in Figure~16 especially in the $\sim 100$~PeV-$1$~EeV range.   

%++++++++++++++++++++++++++++++++++++++++++++++++++++++++++
\begin{figure}[tb]
\begin{center}
\includegraphics[width=0.7\textwidth]{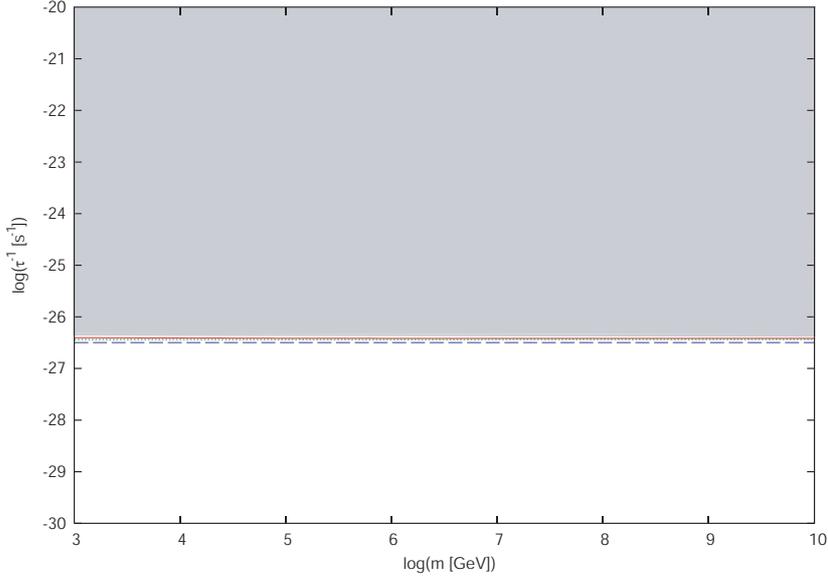}
\end{center}
\caption{The simple cascade gamma-ray bound on the lifetime of decaying VHDM.  The limits are set by the energy budget constraints shown in Figure~1.  The lack of information in this figure is exactly the point: the results are insensitive to dark matter mass and decay channels we consider here.  Note that only the extragalactic contribution is taken into account.  They are consistent with the detailed constraints shown in Figure~16.  
}
\end{figure}
%++++++++++++++++++++++++++++++++++++++++++++++++++++++++++

In Figure~16, both of the extragalactic and Galactic contributions are taken into account.  
The ratio of the Galactic component to the extragalactic component is
\begin{equation}
\frac{\Phi_{\rm G}}{ \Phi_{\rm EG}} \approx \frac{{\mathcal J}_{\Omega} R_{\rm sc} \rho_{\rm sc}}{c t_H  \rho_{\rm dm} \xi_z} \sim 0.5 \frac{{\mathcal J}_{\Omega}}{\xi_z}.
\end{equation} 
For conventional dark matter density profiles such as the NFW profile and Einastro profile~\cite{ein65}, each contribution is typically comparable, though the situation would be affected if the dark matter is more centerally-clustered by e.g., dark matter contraction due to baryon dissipation~\cite{baryon}.  In fact, in Figure~16, the Galactic component comparably contributes to the neutrino constraints.  On the other hand, for the gamma-ray constraints at sufficiently high masses, only the extragalactic component matters since most of the gamma rays do not lie in the LAT range.  
However, when the mass is in the TeV range, the Galactic component in the VHE DGB is more important since the extragalactic component is affected by the EBL attenuation.  The Galactic contribution is also enhanced when dark matter is super heavy (see Figure~16) when UHE electrons/positrons can emit GeV gamma rays via the synchrotron radiation.  
Getting the Galactic part would be relevant for evaluating the dark matter parameters.
Indeed, the case with the Galactic contribution suggests the lower mass, compared to the case without it. 

In Figures~11-14, we showed neutrino spectra as well as gamma-ray spectra that are compatible to the VHE DGB measured by \textit{Fermi}.  As remarked in Section~4.3, the forecasted three-year neutrino constraint will be consistent with the case of ${\rm DM} \rightarrow W^+ W^-$ shown in Figure~11. However, the case of ${\rm DM} \rightarrow \mu^+ \mu^-$ shown in Figure~12 may contradict the neutrino constraint, that is, this case will be tested in a few years.  The cases shown in Figures~13 and 14 are exotic, and they are already excluded by the neutrino observations.   

%++++++++++++++++++++++++++++++++++++++++++++++++++++++++++
\begin{figure*}[bt]
\begin{minipage}{0.49\linewidth}
\begin{center}
\includegraphics[width=\linewidth]{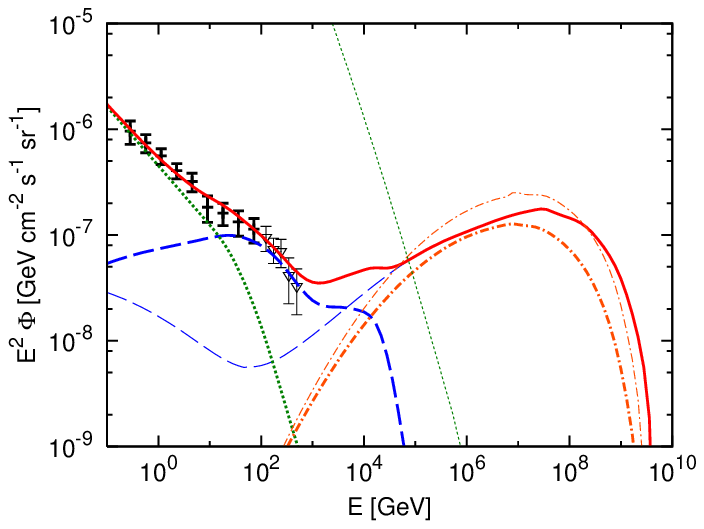}% Here is how to import EPS art
\caption{The same as Figure~11, but the final state spectra are obtained from DGLAP equations.  All the used parameters are the same as those in Figure~14, which indicate that the resulting constraints will be similar to those for the $b \bar{b}$ channel.  For the dark matter component, $m_{\rm dm} c^2=10$~EeV, $\tau_{\rm dm}=4.7 \times {10}^{26}$~s and ${\rm DM} \rightarrow b \bar{b}$ are adopted.  For the power-law component, $a=0.8$ and $\alpha=2.56$ are assumed. This case is ruled out by neutrino observations. 
\newline \,
}
\end{center}
\end{minipage}
\begin{minipage}{.02\linewidth}
\end{minipage}
\begin{minipage}{0.49\linewidth}
\begin{center}
\includegraphics[width=\linewidth]{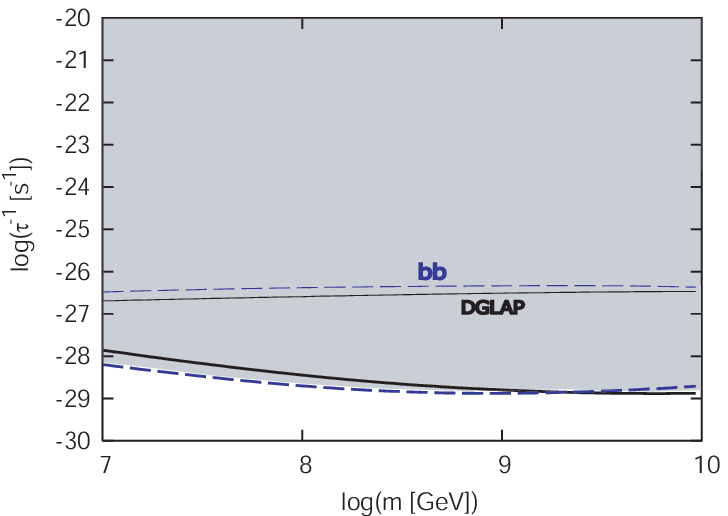}% Here is how to import EPS art
\caption{Influences of the final state spectra obtained from DGLAP equations on neutrino (upper thick curves) and cascade gamma-ray (lower thin curves) constraints.  Decaying VHDM is considered at $m_{\rm dm} c^2 \geq 10$~PeV for  ${\rm DM} \rightarrow q \bar{q}$.  Note that the mass range is different from Figure~16 since we here focus on even higher masses.
\newline \, \newline \, \newline}
\end{center}
\end{minipage}
\end{figure*}
%++++++++++++++++++++++++++++++++++++++++++++++++++++++++++

So far, we have considered final state spectra that are obtained by PYTHIA with extrapolation to higher energies.  For channels with accompanying fragmentation and hadronization, the final state spectra at higher energies become softer at small values of $E/\sqrt{s}$ (i.e., in the low-x regime) because of the scaling violation.  In addition, including supersymmetry (SUSY) leads to much softer spectra~\cite{bs00,bk98,st02}.  In order to demonstrate that our essential results at $\sqrt{s} \gtrsim 10$~PeV are not much changed by such details of the spectra, we show an example for final state spectra obtained by Ref.~\cite{alo+04} (that agrees with Ref.~\cite{st02} within uncertainties), where Dokshitzer-Gribov-Lipatov-Altarelli-Parisi (DGLAP) equations with supersymmetry (SUSY) are solved~\citep[c.f.][]{bs00,bs98,bd03}.  The results are shown in Figure~18, where the spectral shape was calculated for an extreme energy scale $m_{\rm dm} c^2={10}^{14}$~GeV are used (so the realistic spectra would be slightly harder).  The primary spectra in Figure~18 are indeed softer than those in Figure~14, but one sees that all the used parameters are the same as those in Figure~14.  We also compare multi-messenger constraints on the lifetime of decaying VHDM between the above two cases.  As shown in Figure~19, the constraints are similar within about a factor of two.  The reason is that the constraints are not very sensitive to the detailed spectra at small values of $E/\sqrt{s}$.  For the neutrino constraints, in the UHE range, the limits are mostly sensitive to the energy flux around the peak, which does not vary much among various models.  The firmness is enhanced for the gamma-ray constraints.  The cascade gamma-ray bound is connected to the bolometric electromagnetic energy budget, since the initial spectral information is erased and the near-universal spectrum is achieved for sufficiently high masses.  Therefore, only the energy fraction into electromagnetic components of decay products does matter and details of the final state spectrum are irrelevant.  The energy fraction into electromagnetic components is quite similar among the models~\cite{bd03}, so our results shown in Figures~15 and 16 give reasonably accurate results within larger astrophysical uncertainties.

Note that, though we focus on present-day decaying dark matter with $\tau_{\rm dm} > t_H$, there are cosmological bounds, which are more robust but much weaker.  By evaluating how the expansion rate is changed, one can obtain bounds on the lifetime, $\gtrsim {10}^{18}$~s, using CMB data~\cite{cosmodec}.  Though the only assumption is that dark matter particles decay into relativistic particles, it is difficult to improve by further observations, since the decay affects CMB only at large scales and errors are limited by cosmic variance.  Nevertheless, the analyses combined with type Ia supernovae data improve the limit by about an order of magnitude~\cite{cosmodec}.  Also, because of our interests, the mass abundance of relics is fixed to be all of the dark matter.  Allowing for possibilities that $\tau_{\rm dm} \lesssim t_H$ and/or decaying particles are subdominant, one can obtain more general constraints on relic neutrals from the Big-Bang nucleosynthesis as well as the diffuse backgrounds~\citep[e.g.,][]{ell+92,gon+93,kr97,ell+85}.      

%%%%%%%%%%%%%%%%%%%%%%%%%%%%%%%%%%%%%%%%%%%%%%%%%%%%%%%%
%%%%%%%%%%%%%%%%%%%%%%%%%%%%%%%%%%%%%%%%%%%%%%%%%%%%%%%%

\section{Summary and discussion}

In this paper, we investigate diffuse neutrino and gamma-ray backgrounds coming from VHDM with $m_{\rm dm} c^2 \gtrsim$~TeV, on the basis of observations by \textit{Fermi} and IceCube.  Our results are summarized as follows. 
 
(1) We pointed out the possibility that VHDM could contribute to the VHE DGB.  
This result is based on the recent indication of the ``VHE Excess" that is identified in the \textit{Fermi} era from the fact that the simple extrapolation of the astrophysical power-law component falls well below the measured DGB due to the severe EBL attenuation.  We mainly considered decaying VHDM as the possible scenario, since quite large values of the cross section and the boost factor are required for annihilating VHDM.  

Although the precise determination of the VHE DGB is obviously required and other interpretations are also possible, significant impacts on understanding dark matter and the DGB are expected if the VHE Excess is confirmed.  Such the optimistic scenario is also interesting because they are testable for some parameters.  In the Galactic scenario, dark matter masses of $m_{\rm dm} c^2 \sim 3-30$~TeV are necessary to explain the VHE DGB.  Although the neutrino background is below the atmospheric background, the case of leptonic channels is testable by IceCube within a few years.  On the other hand, in the VHDM-induced cascade scenario, extremely heavy masses ($\gtrsim 10$~PeV) are required.  Since large neutrino fluxes are predicted, such exotic cases are already excluded by neutrino observations.  Those results demonstrate the importance of neutrino detectors such as IceCube and KM3Net for testing the properties of dark matter.  Of course, deeper observations by \textit{Fermi} and future detectors such as GAMMA-400~\cite{GAM} should also provide us with crucial information.

Although we considered the isotropic backgrounds for conservative studies, future more detailed analyses using the angular information can give tighter constraints.  
The dipole anisotropy can be expected in the Galactic scenario, so searches for anisotropy in the VHE range would also be useful.  

In addition, we also expect that not only neutrinos but also TeV gamma-ray observations via Cherenkov telescopes such as HAWC and CTA will be helpful.  If high-energy gamma rays are produced above TeV energies due to annihilation or decay of VHDM, there may be a bump around 1~TeV or higher energies in the gamma-ray spectrum of galaxies.  Predicted fluxes depend on dark matter parameters, but future searches for TeV gamma rays from nearby galaxies such as LMC and M31 will be relevant.  

(2) We revisited the cosmic energy budget argument in light of \textit{Fermi} and IceCube observations, and applied it to VHDM.  In the high mass range, gamma-ray cascades must be taken into account for the extragalactic contribution.  We performed detailed calculations of both neutrinos and gamma rays, which enables us to demonstrate the power of neutrino observations compared to gamma-ray observations. 

We derived a new cascade gamma-ray bound both analytically and numerically.  In the sufficiently high-mass range, this cascade gamma-ray bound gives conservative constraints, so that it can be used reasonably independently of channels and it is also useful to restrict the total decay rate and the total annihilation cross section of VHDM in the high mass range.  Importantly, since the \textit{Fermi} DGB is lower than the EGRET one, the constraints are improved by about a factor $\sim 10$.  In particular, for decaying VHDM, we obtained $\tau_{\rm dm} \gtrsim {10}^{26}-{10}^{27}$~s, which is more stringent than the neutrino limits at $\lesssim 10-100$~TeV.  On the other hand, at higher mass scales, the neutrino limits are stronger, while the DGB gives more conservative constraints.  The cascade gamma-ray bound is largely mass-independent at sufficiently high masses.  For annihilating VHDM, the gamma-ray limits are typically stronger in the relevant mass range where the unitarity bound is preserved.  But neutrino observations give more conservative but still interesting constraints compared to the unitarity bound.   

We focused on gamma-ray and neutrino constraints in light of \textit{Fermi} and IceCube/KM3Net, but cosmic-ray experiments including observations of UHE gamma rays and the arrival distribution in UHE cosmic rays are also useful in the wimpzillas regime.  The importance of such UHE particle observations was studied in early days, and top down models for UHE cosmic rays have been constrained.   

%%%%%%%%%%%%%%%%%%%%%%%%%%%%%%%%%%%%%%%%%%%%%%%%%%%%%%%%
%%%%%%%%%%%%%%%%%%%%%%%%%%%%%%%%%%%%%%%%%%%%%%%%%%%%%%%%

\section*{Acknowledgments}
K.~M. is supported by JSPS and CCAPP. 
The research of J.~F.~B. is supported by NSF Grant PHY-1101216.  
We thank Shunsaku Horiuchi, Boaz Katz, Alexander Kusenko, Ranjan Laha, and especially Carsten Rott for helpful discussions. 

%%%%%%%%%%%%%%%%%%%%%%%%%%%%%%%%%%%%%%%%%%%%%%%%%%%%%%%%
%%%%%%%%%%%%%%%%%%%%%%%%%%%%%%%%%%%%%%%%%%%%%%%%%%%%%%%%

\appendix
\section{Intergalactic cascades}

Taking into account the pair creation, inverse Compton, synchrotron radiation and adiabatic loss, we numerically calculate the cascade emission by solving the Boltzmann equations that are often referred as kinetic equations~\cite{bs00,lee98,mur12},   
\begin{eqnarray}
\frac{\pd N_\gamma}{\pd x} &=& -N_{\gamma} R_{\gamma \gamma} 
+ \frac{\pd N_{\gamma}^{\rm IC}}{\pd x} 
+ \frac{\pd N_{\gamma}^{\rm syn}}{\pd x}
- \frac{\pd}{\pd E} [P_{\rm ad} N_e]  
+ Q_{\gamma}^{\rm inj},\\
\frac{\pd N_e}{\pd x} &=& \frac{\pd N_e^{\gamma \gamma}}{\pd x} 
-N_{e} R_{\rm IC} + \frac{\pd N_e^{\rm IC}}{\pd x} 
- \frac{\pd}{\pd E} [(P_{\rm syn}+P_{\rm ad}) N_e] 
+ Q_e^{\rm inj}, 
\end{eqnarray}
where 
\begin{eqnarray}
R_{\gamma \gamma} &=& \int d \varepsilon \frac{dn }{d \varepsilon}  \int \frac{d \Omega}{4 \pi} \,\, \tilde{c} \sigma_{\gamma \gamma} (\varepsilon, \Omega), \nonumber \\
R_{\rm IC} &=& \int d \varepsilon \frac{dn }{d \varepsilon} \int \frac{d \Omega}{4 \pi} \,\, \tilde{c}  \sigma_{\rm IC} (\varepsilon, \Omega), \nonumber \\
\frac{\pd N_\gamma^{\rm IC}}{\pd x} &=& \int d E' N_{e} (E') \,\, \int d \varepsilon \frac{dn }{d \varepsilon} \,\, \int  \frac{d \Omega}{4 \pi} \,\, \tilde{c}   \frac{d \sigma_{\rm IC}}{d E_\gamma} (\varepsilon, \Omega, E'), \nonumber \\
\frac{\pd N_e^{\gamma \gamma}}{\pd x} &=& \int d E' N_{\gamma} (E') \,\, \int d \varepsilon \frac{dn }{d \varepsilon} \,\, \int \frac{d \Omega}{4 \pi} \,\, \tilde{c}  \frac{d \sigma_{\gamma \gamma}}{d E_e} (\varepsilon, \Omega, E'), \nonumber \\
\frac{\pd N_e^{\rm IC}}{\pd x} &=& \int d E' N_{e} (E') \,\, \int d \varepsilon \frac{dn }{d \varepsilon}  \,\, \int \frac{d \Omega}{4 \pi} \,\, \tilde{c}  \frac{d \sigma_{\rm IC}}{d E_e} (\varepsilon, \Omega, E'). 
\end{eqnarray}
Here $\tilde{c} =(1-\mu) c$, $P_{\rm syn}$ is the synchrotron energy loss rate, $P_{\rm ad}$ is the adiabatic energy loss rate, $N_\gamma$ and $N_e$ are photon and electron/positron number densities per energy decade, and $Q_\gamma^{\rm inj}$ and $Q_e^{\rm inj}$ are photon and electron/positron injection rate.   

For the intergalactic cascade, $x$ corresponds to travel distance (divided by $c$), whereas time is used as $x$ for the source cascade. This paper focuses on the former case, so we consider the CMB and EBL as a target photon spectrum, $d n/d \varepsilon$.  As an EBL model, we use the low-IR model of Ref.~\cite{kne+04}, which gives relatively conservative estimates.  For variations in EBL models, we refer to Ref.~\cite{fin+10}.  The highly uncertain cosmic radio background is not taken into account in this work, but it is relevant for direct detections of UHE gamma rays themselves~\cite{mur09,mur12}. 
Relevant length-scales are depicted in Figure~20. 

%++++++++++++++++++++++++++++++++++++++++++++++++++++++++++
\begin{figure*}[bt]
\begin{minipage}{0.49\linewidth}
\begin{center}
\includegraphics[width=\linewidth]{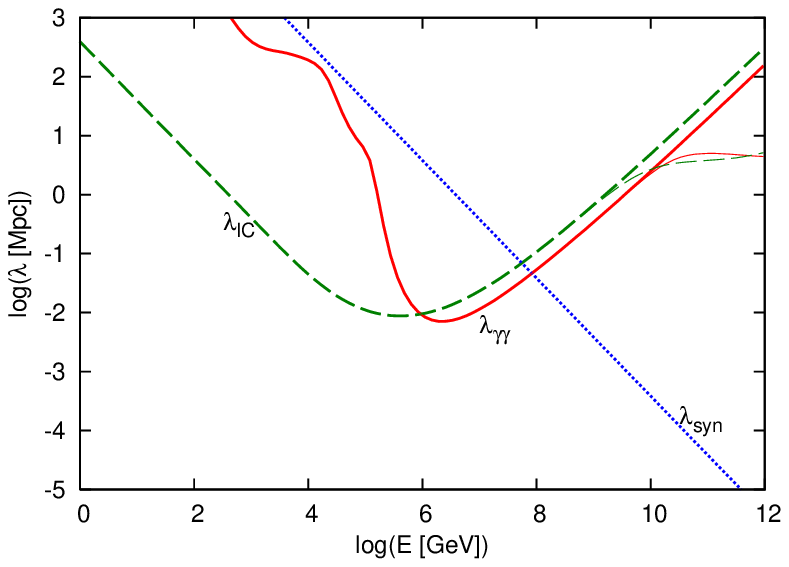}% Here is how to import EPS art
\caption{Interaction length (solid curve) of high-energy photons for the pair creation process, the energy loss length of electron-positron pairs for the inverse Compton process (dashed curve), and the synchrotron cooling length for ${10}^{1.5}$~nG that can be expected in structured regions of the universe (dotted curve).  Thick/thin curves represent cases without/with the CRB.
}
\end{center}
\end{minipage}
\begin{minipage}{.02\linewidth}
\end{minipage}
\begin{minipage}{0.49\linewidth}
\begin{center}
\includegraphics[width=\linewidth]{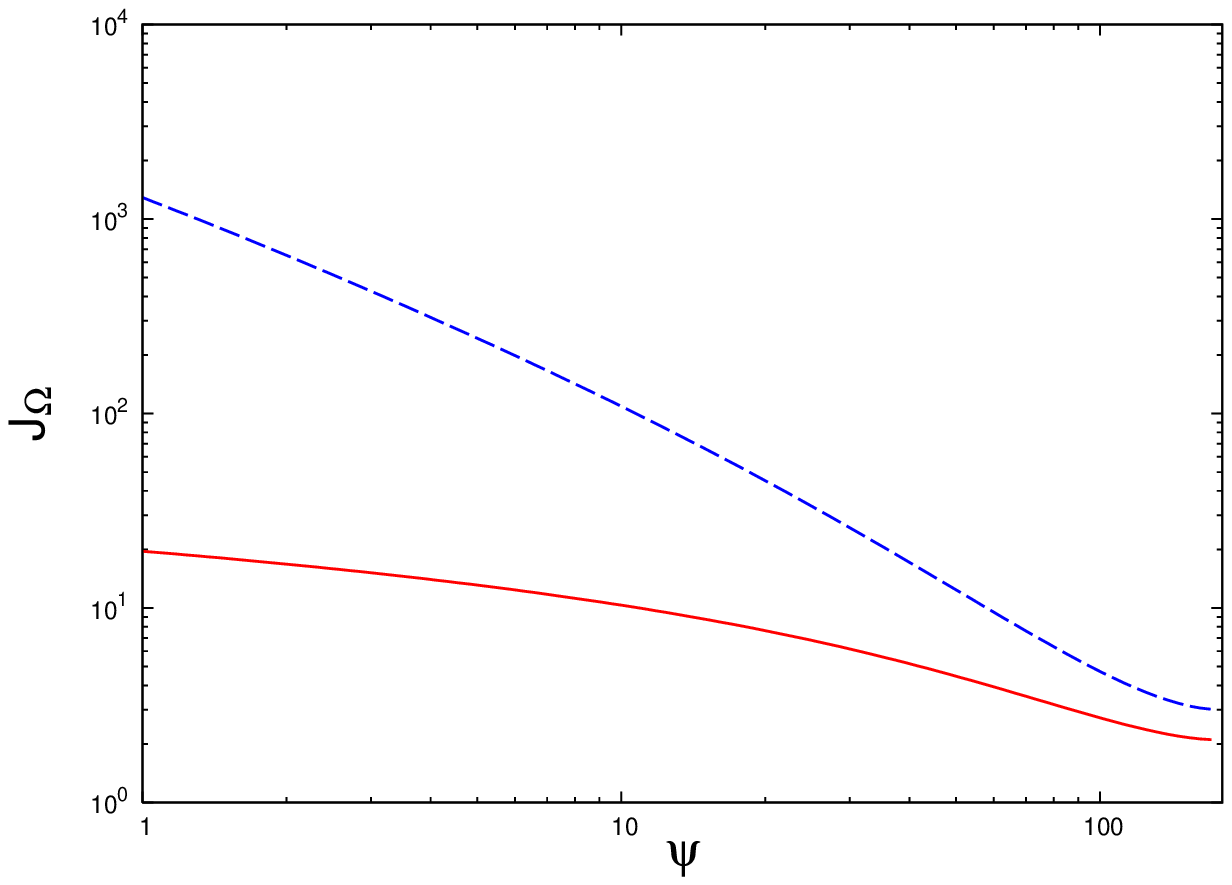}% Here is how to import EPS art
\caption{The average of the line of sight integration, ${\mathcal J}_{\Omega}$, inside a cone with half angle $\psi$ around the Galactic center, as a function of the pointing angle with respect to the Galactic center direction for the NFW profile.  The solid curve is for decaying dark matter, whereas the dashed curve is for annihilating dark matter.
\newline
}
\end{center}
\end{minipage}
\end{figure*}
%++++++++++++++++++++++++++++++++++++++++++++++++++++++++++

%%%%%%%%%%%%%%%%%%%%%%%%%%%%%%%%%%%%%%%%%%%%%%%%%%%%%%%%
%%%%%%%%%%%%%%%%%%%%%%%%%%%%%%%%%%%%%%%%%%%%%%%%%%%%%%%%

\section{Galactic contribution}

In this section, we describe how to calculate the Galactic contribution of dark matter. 
As for spectra of primary particles, the Galactic contribution is given by
\begin{eqnarray}
I_E^{\rm dec} (\psi) &=& \frac{1}{4 \pi} \int dl \, \, \, \frac{\rho_{\rm dm} (r)}{m_{\rm dm} \tau_{\rm dm}}  \frac{d S}{d E} 
= \frac{1}{4 \pi m_{\rm dm} \tau_{\rm dm}} \frac{d S}{dE} \int_0^{l_{\rm max}} dl \,\,\, \rho_{\rm dm} (r) \nonumber \\
&=& \frac{R_{\rm sc} \rho_{\rm sc}}{4 \pi m_{\rm dm} \tau_{\rm dm}} \frac{d S}{dE}  {\mathcal J}^{\rm dec} (\psi)
\end{eqnarray}
for decaying dark matter, and 
\begin{eqnarray}
I_E^{\rm ann} (\psi) &=& \frac{1}{4 \pi} \int dl \,\,\, \frac{{<\sigma v>}_{\rm dm}}{2} {\left( \frac{\rho_{\rm dm} (r)}{m_{\rm dm}} \right)}^2  \frac{d S}{d E}  
= \frac{{<\sigma v>}_{\rm dm}}{8 \pi m_{\rm dm}^2}  \frac{d S}{d E} \int_0^{l_{\rm max}} dl \,\,\, \rho_{\rm dm}^2 (r) \nonumber \\
&=& \frac{{<\sigma v>}_{\rm dm} R_{\rm sc} \rho_{\rm sc}^2}{8 \pi m_{\rm dm}^2}  \frac{d S}{d E}{\mathcal J}^{\rm ann} (\psi)
\end{eqnarray}
for annihilating dark matter. 

We have introduced the $\mathcal J$ factor.  For decaying dark matter, one has
\begin{equation}
{\mathcal J}^{\rm dec} (\psi)= \frac{1}{R_{\rm sc} \rho_{\rm sc}} \int_0^{l_{\rm max}} dl \,\,\, \rho_{\rm dm} (r). 
\end{equation}
For annihilating dark matter, one has
\begin{equation}
{\mathcal J}^{\rm ann} (\psi) = \frac{1}{R_{\rm sc} \rho_{\rm sc}^2} \int_0^{l_{\rm max}} dl \,\,\, \rho_{\rm dm}^2 (r) 
\end{equation}
Here $r=\sqrt{R_{\rm sc}^2-2 l R_{\rm sc} \cos \psi +l^2}$, $l_{\rm max}=R_{\rm sc} \cos \psi + \sqrt{R_{\rm mw}^2-R_{\rm sc}^2 \sin^2 \psi}$, and $R_{\rm mw}=20$~kpc is the size of our Galaxy.  We take $R_{\rm sc}=8.5$~kpc and $\rho_{\rm sc} c^2=0.3~{\rm GeV}~{\rm cm}^{-3}$ with the NFW profile.  
Note that ${\mathcal J} (\psi)$ never vanishes, and the minimal isotropic component is written as ${\mathcal J} (180^\circ)$.  By evaluating Eqs.~(B.3) and (B.4), we obtain $\simeq 1$ for the decay and $\simeq 0.4$ for the annihilation~\cite{yuk+07}, respectively.  It is often useful to see the average of ${\mathcal J}$ in a cone with half-angle $\psi$ around the Galactic center, which is defined as
\begin{eqnarray}
{\mathcal J}_{\Omega} = \frac{2 \pi}{\Omega} \int_{\cos \psi}^1 d (\cos \psi') \,\,\, {\mathcal J} (\psi')
\end{eqnarray}
where $\Omega=2 \pi (1-\cos \psi)$ is the solid angle of a field of view, and ${\mathcal J}_\Omega$ is shown in Figure~21.

For the Galactic component, the mean free path of gamma rays is $\gtrsim 10$~kpc, so that the primary gamma-ray-induced cascade would not be much critical.  On the other hand, electrons and positirons are largely confined and should lose their energies as radiation, so that 
resulting secondary emissions have to be considered.  Instead of performing time-dependent calculations, in this work, we obtain the stead-state distribution. In the continuous energy-loss approximation, the kinetic equation becomes 
\begin{equation}
\frac{\pd N_e}{\pd t} = -\frac{\pd}{\pd E} [(P_{\rm IC}+P_{\rm syn}) N_e] + Q_e^{\rm inj}, 
\end{equation}
where $P_{\rm IC}$ is the inverse Compton energy loss rate.  Hence, the steady state distribution is easily obtained by
\begin{equation}
N_e =  \frac{1}{P_{\rm IC}+P_{\rm syn}} \int_E d E' \,\,\, Q_e^{\rm inj} 
\end{equation}
For a given electron/positron distribution, the gamma-ray spectrum coming from synchrotron and inverse-Compton emission is evaluated simultaneously from Eq.~(A.3).  We adopt $1~\mu$G as the Galactic magnetic field.  

%%%%%%%%%%%%%%%%%%%%%%%%%%%%%%%%%%%%%%%%%%%%%%%%%%%%%%%%%%%%%%%%%%%%%%%%%%%%%%%%%%%%%%
%%%%%%%%%%%%%%%%%%%%%%%%%%%%%%%%%%%%%%%%%%%%%%%%%%%%%%%%%%%%%%%%%%%%%%%%%%%%%%%%%%%%%%
\newpage

\bibliographystyle{JHEP}
\bibliography{ms}

%\begin{thebibliography}{}
%\end{thebibliography}
%\bibliographystyle{JHEP}

\end{document}